\documentclass[manuscript]{acmart}

\usepackage{algorithm}
\usepackage{algorithmic}
\usepackage{subfigure}
\usepackage{multirow}
\usepackage{tabularx}
\usepackage{caption}

\usepackage{amssymb}
\usepackage{bbding}
\usepackage{paralist}

\AtBeginDocument{%
  }

\setcopyright{none}
\acmDOI{1234567.1234567}

\renewcommand\footnotetextcopyrightpermission[1]{} % removes footnote with conference information in the first column
\begin{document}
% \captionsetup[table]{skip=3pt}
% \captionsetup[figure]{skip=3pt}
%%
%% The "title" command has an optional parameter,
%% allowing the author to define a "short title" to be used in page headers.
% \title{AQP: Accelerating Quadratic Programming in Path Planning with High Data Sparsity}

%% \title{Aplus: Accelerating Autonomous Path Planning on FPGAs with Sparsity-Aware HW/SW Co-Optimizations}
% \title{A Sparsity-Aware Autonomous Path Planning Accelerator with Algorithm-Architecture Co-Design}
\title{A Sparsity-Aware Autonomous Path Planning Accelerator with HW/SW Co-Design and Multi-Level Dataflow Optimization}

%% A Sparsity-Aware Autonomous Path Planning Accelerator with Algorithm-Architecture Co-Design

% %%
% %% The "author" command and its associated commands are used to define
% %% the authors and their affiliations.
% %% Of note is the shared affiliation of the first two authors, and the
% %% "authornote" and "authornotemark" commands
% %% used to denote shared contribution to the research.
%\author{Ben Trovato}
%\authornote{Both authors contributed equally to this research.}
%\email{trovato@corporation.com}
%\orcid{1234-5678-9012}

%\author[Y. Zhang, X. Niu, Y. Zhang, H. Tian, B. Yu, S. Liu, S. Huang]{\Large Yanjun Zhang\textsuperscript{*1},
%Xiaoyu Niu\textsuperscript{*1},
%Yifan Zhang\textsuperscript{2},
%Hongzheng Tian\textsuperscript{2},
%Bo Yu$^\dagger$\textsuperscript{3},
%Shaoshan Liu\textsuperscript{3},
%Sitao Huang$^\dagger$\textsuperscript{2}
%}

%\affiliation{%
%\institution{\textsuperscript{1}Beijing Institute of Technology,
%\textsuperscript{2}University of California, Irvine,\\
%\textsuperscript{3}Shenzhen Institute of Artificial Intelligence and Robotics for Society}
%\authornote{indicates equal contribution to the paper. $^\dagger$ indicates corresponding author.}
%}

\thanks{This manuscript is an extension of a conference paper. The previously accepted conference paper is titled \emph{A Sparsity-Aware Autonomous Path Planning Accelerator with Algorithm-Architecture Co-Design}, to be published in 2024 ACM/IEEE International Conference on Computer-Aided Design (ICCAD). The previous paper only focus on accelerating a single module in path planning flow and proposing efficient hardware for individual operators. This paper provides new optimizations from a system perspective. We analyze inter-operator data dependencies, propose an operator fusion scheme to overlap the latency. The proposed operator fusion method achieves a fine-grained pipeline across operators that significantly reduces latency with negligible overhead. It also saves unnecessary memory access and logic resources. At the system level, we map different steps of the planning process to the CPU and FPGA and pipeline these steps to enhance end-to-end throughput. This paper also performs a knowledge-based search for optimal parameters to accelerate the algorithm convergence. Compared with previous conference paper, our work acheives 1.48x latency speedup and 2x end-to-end throughput.
}

\author{Yifan Zhang}
\affiliation{%
 \institution{University of California, Irvine}
 \country{USA}}
\email{yifanz58@uci.edu}

\author{Xiaoyu Niu}
\affiliation{%
 \institution{Beijing Institute of Technology}
 \country{China}}
\author{Hongzheng Tian}
\affiliation{%
 \institution{University of California, Irvine}
 \country{USA}}

\author{Yanjun Zhang}
\affiliation{%
 \institution{Beijing Institute of Technology}
 \country{China}}

\author{Bo Yu}
\authornote{Corresponding authors.}
\affiliation{%
 \institution{Shenzhen Institute of Artificial Intelligence and Robotics for Society}
 \country{China}}

\author{Shaoshan Liu}
\affiliation{%
 \institution{Shenzhen Institute of Artificial Intelligence and Robotics for Society}
 \country{China}}

\author{Sitao Huang}
\authornotemark[1]
% \footnote{$*$ indicates corresponding authors.}
\affiliation{%
 \institution{University of California, Irvine}
 \country{USA}}
\email{sitaoh@uci.edu}

\begin{abstract}
% Path planning is a critical task in autonomous driving systems that is most susceptible to real-time constraints but often demands computationally intensive mathematical solvers, two contradictory goals. This conflict makes the computing of path planning a paramount challenge. At the heart of most path planners is the quadratic programming (QP) solver, which places excessive demands on the CPU in real-world autonomous driving applications. In this paper, we present an FPGA-based acceleration framework for path planning problems. Our approach leverages an operator splitting solver for quadratic programs (OSQP) and employs the preconditioned conjugate gradient (PCG) method for solving linear systems, which are customized to be more hardware-friendly than prior works. Specific memory management and parallel processing were tailored to the matrix pattern, and the incorporation of pipelining was executed to enhance throughput and execution speed. Our FPGA-based implementation achieves state-of-the-art performance against existing works, including an average 1.98$\times$ speedup compared with the state-of-the-art QP solver on Intel i7-11800H CPU, 3.90$\times$ speedup over an ARM Cortex-A57 embedded CPU, and 12.3$\times$ speedup over an NVIDIA RTX 3090 GPU.
Path planning is a critical task for autonomous driving, aiming to generate smooth, collision-free, and feasible paths based on input perception and localization information. The planning task is both highly time-sensitive and computationally intensive, posing significant challenges to resource-constrained autonomous driving hardware. In this paper, we propose an end-to-end framework for accelerating path planning on FPGA platforms. This framework focuses on accelerating quadratic programming (QP) solving, which is the core of optimization-based path planning and has the most computationally-intensive workloads. Our method leverages a hardware-friendly alternating direction method of multipliers (ADMM) to solve QP problems while employing a highly parallelizable preconditioned conjugate gradient (PCG) method for solving the associated linear systems. We analyze the sparse patterns of matrix operations in QP and design customized storage schemes along with efficient sparse matrix multiplication and sparse matrix-vector multiplication units. Our customized design significantly reduces resource consumption for data storage and computation while dramatically speeding up matrix operations. Additionally, we propose a multi-level dataflow optimization strategy. Within individual operators, we achieve acceleration through parallelization and pipelining. For different operators in an algorithm, we analyze inter-operator data dependencies to enable fine-grained pipelining. At the system level, we map different steps of the planning process to the CPU and FPGA and pipeline these steps to enhance end-to-end throughput. We implement and validate our design on the AMD ZCU102 platform. Our implementation achieves state-of-the-art performance in both latency and energy efficiency compared to existing works, including an average 1.48$\times$ speedup over the best FPGA-based design, a 2.89$\times$ speedup compared to the state-of-the-art QP solver on an Intel i7-11800H CPU, a 5.62$\times$ speedup over an ARM Cortex-A57 embedded CPU, and a 1.56$\times$ speedup over state-of-the-art GPU-based work. Furthermore, our design delivers a 2.05$\times$ improvement in throughput compared to the state-of-the-art FPGA-based design.

\end{abstract}


%%
%% Keywords. The author(s) should pick words that accurately describe
%% the work being presented. Separate the keywords with commas.
%% A "teaser" image appears between the author and affiliation
%% information and the body of the document, and typically spans the
%% page.
% \begin{teaserfigure}
%   \includegraphics[width=\textwidth]{sampleteaser}
%   \caption{Seattle Mariners at Spring Training, 2010.}
%   \Description{Enjoying the baseball game from the third-base
%   seats. Ichiro Suzuki preparing to bat.}
%   \label{fig:teaser}
% \end{teaserfigure}

% \received{20 February 2007}
% \received[revised]{12 March 2009}
% \received[accepted]{5 June 2009}

%%
%% This command processes the author and affiliation and title
%% information and builds the first part of the formatted document.
\maketitle

\section{Introduction}

In the realm of autonomous driving, the capacity for swift and accurate path planning is crucial, serving as a critical component of the vehicle's computing pipeline ~\cite{liu2018creating,liu2019edge,hao2024orianna}. Since it is the most computationally expensive module at the backend of the autonomous driving pipeline, it is also the most susceptible to real-time constraints. The path planning process not only dictates the feasibility, reliability, and safety of the whole autonomous driving system but also directly influences the responsiveness and adaptability, and hence safety, of the vehicle in dynamic environments ~\cite{wan2024vpp,yu2020building}. 

Traditional computational approaches towards planning often confront dual challenges of meeting real-time processing requirements and managing the complex, data-intensive computations needed for effective path planning ~\cite{apolo,lavalle2006planning,wan2021survey}. These challenges underscore the pressing need for innovative solutions that can deliver both speed and accuracy. To address these demands, this paper introduces a novel FPGA-based acceleration framework for enhancing the path planning capabilities of autonomous vehicles. 

\begin{figure}
    \centering
    \includegraphics[width=\linewidth]{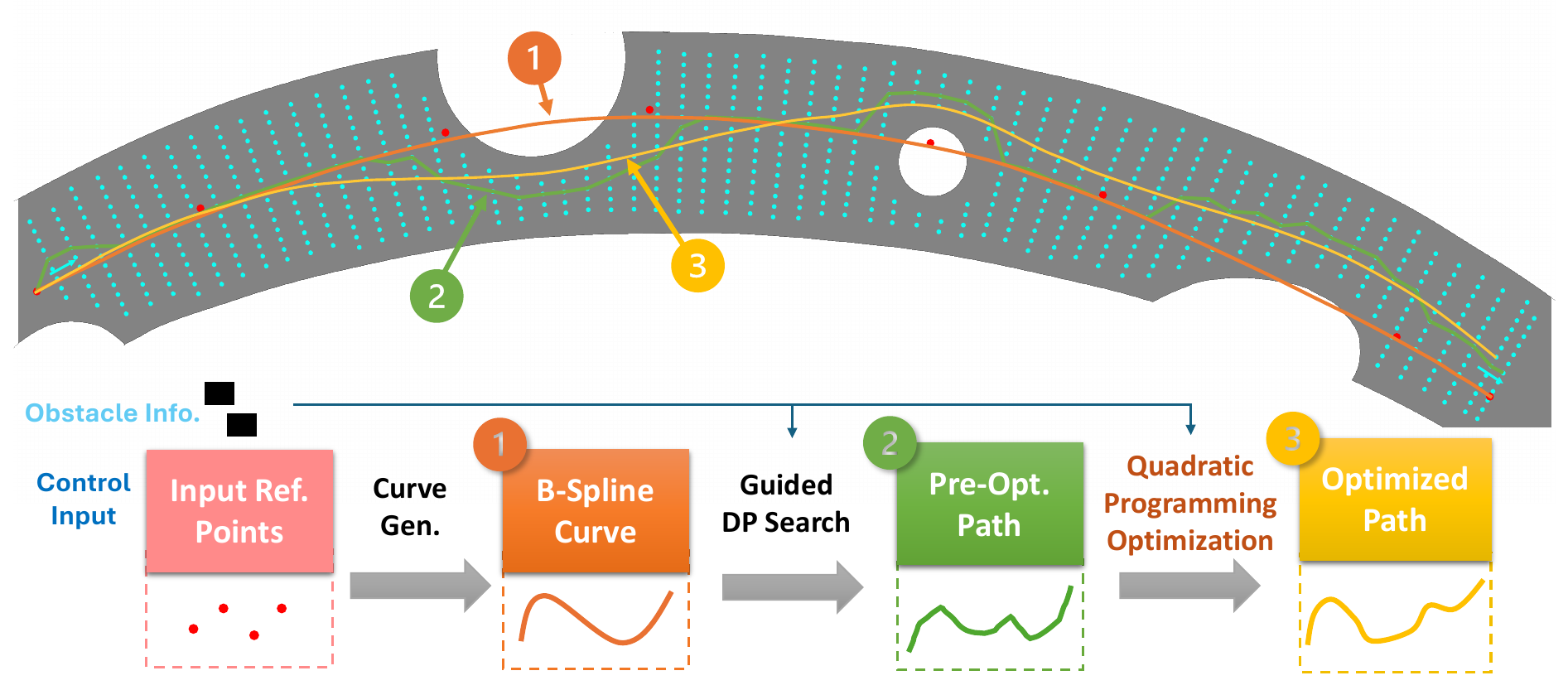}
    \caption{Overview of the Path Planning Flow}
    \Description{This figure shows a path planning example and the three stages of our path planning framework. }
    \vspace{-2em}
    \label{fig:framework}
\end{figure}

In detail, a path planner typically starts with a given \emph{global path}, which connects the start point and the destination point with a rough curve without detailed kinematical considerations. Starting with this global path, the path planner goes through a number of path refinement iterations where that path is adjusted so that it optimizes the objective function while meeting all the constraints. In this process, the path is smoothened, and all mechanical and kinematical constraints are considered. Figure~\ref{fig:framework} shows an example of path planning. 

%% current path planning solutions and their limitations

%Existing path planning methods could be categorized into \emph{search}-based, \emph{sampling}-based, and \emph{optimization}-based method. Search-based methods use algorithms like A* and Dijkstra to search for the optimal or near-optimal path that meets the constraints. Due to the huge search space, search-based methods are computationally expensive. Besides, the quality of the generated paths may not meet the requirements. Sampling-based methods generate candidate paths from certain probabilistic distribution or assumptions, and select the paths according to the pre-defined optimization objectives. Similar to search-based methods, sampling-based methods also suffer from long search time in the large design space. Optimization-based method formulates the path planning problem as an optimization problem and solve it with general optimization solvers, which produces the optimal path for the chosen objective function. Even though the optimization-based methods find high-quality optimal paths, it takes longer time to find the solution. 

The major limitations of commercial path planning solutions come from several different aspects ~\cite{li2020autonomous}. First, at the algorithm level, faster, easier-to-compute, greedy methods compromise on the planning quality. It is challenging to achieve a good balance between path planning quality and computation time. Second, at the libraries and tools level, current solutions typically rely on off-the-shelf general linear algebra and optimization libraries and tools, which are designed for general problems. These general library calls ignore domain-specific (path planning) information and heuristics and therefore do not fully leverage domain-specific optimization opportunities. Third, at hardware and system level, existing solutions usually assume general-purpose processing systems as the underlying computing platform, and ignore the differences between different platforms and opportunities from system customization. 

%As researchers and developers push for more advanced level of autonomous driving, we are facing higher requirements on path quality, more constraints, and rapidly changing dynamic environment. The complexity in the existing path planning solutions poses prohibitive computing cost for advanced autonomous driving tasks. Clearly, to mitigate this challenge, we need sophisticated path planning algorithms that can produce high-quality path, while at the same time they should achieve  high computational efficiency via deep algorithm-architecture customization. 

%% our proposed methods and their advantage, brief
%In this work, we propose \emph{Aplus}, an efficient path planning acceleration framework designed for autonomous vehicles. The accelerator leverages task and platform specific information, including sparsity, data bitwidth, computational intensity, etc., and automatically generates sparsity-aware hardware-software co-optimized hardware acceleration systems. The major contributions of this work can be summarized as follows. 
%ICCAD VERSION:
% In this work, we propose an efficient path planning acceleration framework designed for autonomous vehicles. The accelerator leverages task and platform specific information, including sparsity, data bitwidth, computational intensity, etc., and automatically generates sparsity-aware hardware-software co-optimized hardware acceleration systems. The major contributions of this work can be summarized as follows.
The proposed framework leverages task and platform-specific information, including sparsity, problem size, datatype, computational intensity of each module, etc., to design efficient sparsity-aware storage schemes and computing units. We also proposed a fine-grained multi-level dataflow optimization to maximize end-to-end performance. The major contributions of this work can be summarized as follows.
%: 1) We propose an end-to-end FPGA accelerated path planning framework designed for autonomous vehicles, which features sparsity-aware hardware-software co-optimizations; 2) We propose algorithm and hardware customizations for the alternating direction method of multipliers (ADMM)~\cite{ADMM} based quadratic programming solver, which leverages planning task specific structured optimization opportunities and effectively speeds up path planning with fine-grained hardware acceleration. 3) On real-world collected and simulated datasets, our proposed framework achieves on average 1.98$\times$ speedup compared with the Intel i7 CPU, and 3.90$\times$ speedup compared with the ARM Cortex-A57.

\begin{enumerate}
    %\item We propose \emph{Aplus}, an end-to-end FPGA accelerated path planning framework designed for autonomous vehicles, which features sparsity-aware hardware-software co-optimizations. 
    %ICCAD version:
    % \item We propose an end-to-end FPGA accelerated path planning framework designed for autonomous vehicles, which features sparsity-aware hardware-software co-optimizations. 
    \item We propose an end-to-end energy-efficient FPGA-accelerated path planning framework for autonomous vehicles, which features sparsity-aware hardware-software co-optimizations with multi-level dataflow optimizations.
    %\item We propose algorithm and hardware customizations for the alternating direction method of multipliers (ADMM) \cite{ADMM} based quadratic programming solver inside Aplus, which leverages planning task specific structured optimization opportunities and effectively speeds up path planning with fine-grained hardware acceleration. 
    %ICCAD version
    % \item We propose algorithm and hardware customizations for the alternating direction method of multipliers (ADMM)~\cite{ADMM} based quadratic programming solver, which leverages planning task specific structured optimization opportunities and effectively speeds up path planning with fine-grained hardware acceleration. 
    \item We propose hardware customizations for the alternating direction method of multipliers (ADMM)~\cite{ADMM} based quadratic programming solver, which leverages planning task specific structured sparsity to effectively speed up path planning with fine-grained parallelization and pipelining. 
    %\item On real-world collected and simulated datasets, our proposed Aplus framework achieves on average 2.94$\times$ speedup compared with the Intel i7 CPU, 5.03$\times$ speedup compared with the ARM Cortex-A57.
    \item We propose a multi-level dataflow optimization strategy to maximize end-to-end performance with negligible resource overhead. At the inter-operator level, we analyze data dependencies across different operators in the algorithm and propose fine-grained inter-operator pipelining. At the system level, we map different independent stages of the planning process into several threads on the CPU and FPGA and enable pipelining for those threads to enhance end-to-end throughput.
    \item We perform a design space exploration for algorithm-architecture co-optimizations. We search for the proper setting of the fix-point datatype for optimal hardware efficiency without compromising numerical precision. We also find the optimal algorithm parameters for fast convergence.

    \item On real-world collected and simulated datasets, our proposed framework achieves on average 1.48$\times$ speedup over the state-of-the-art FPGA-based design, a 2.89$\times$ speedup compared to the state-of-the-art QP solver on an Intel i7-11800H CPU, a 5.62$\times$ speedup over an ARM Cortex-A57 embedded CPU, and a 1.56$\times$ speedup over the state-of-the-art GPU-based implementation on NVIDIA RTX 3090 GPU.
\end{enumerate}

% The rest of this paper is organized as follows. Section~\ref{sec:related} reviews the existing path planning solutions. Section~\ref{sec:framework} formulates the path planning problem and discusses algorithm-level challenges and opportunities. We introduce the designed framework and algorithm-architecture co-optimizations in Section~\ref{sec:aplus} and ~\ref{sec:path_planner_opt}, and show evaluation results in Section~\ref{sec:results}. Finally, we conclude the paper in Section~\ref{sec:conclusion}. 

\section{Related Works}\label{sec:related}

\subsection{Path Planning for Autonomous Driving}
In an autonomous driving system, the path planning subsystem utilizes data on obstacle positions and shapes from the perception module to generate a collision-free, smooth and dynamically feasible path for the control module, accounting for the vehicle's kinematic constraints.
However, finding the optimal path in complex traffic scenarios presents a significant challenge. This is due to the vast search space encompassing possible vehicle configurations (positions and headings) and the need for real-time decision-making. 
Traditional path planning algorithms address this challenge with a two-stage process~\cite{fan2018baidu,gasparetto2015path}: path finding and trajectory optimization. Path finding focuses on identifying a collision-free path within the configuration space, laying the groundwork for the subsequent path-smoothing stage. Trajectory optimization then refines this path, focusing on smoothness while ensuring continued obstacle avoidance and adherence to the vehicle's dynamic constraints.

Researchers have developed several search-based path-finding algorithms. These algorithms discretize the configuration space into grid structures and employ efficient shortest path finding techniques, such as the hybrid A$^*$ algorithm~\cite{dolgov2008practical,zhong2020hybrid}, to identify the shortest solution. Search-based algorithms struggle with high computational complexity when dealing with large-scale planning problems. To address this limitation, sampling-based algorithms, such as Rapidly-exploring Random Trees (RRT) and its variants~\cite{bruce2002real,karaman2011anytime,noreen2016optimal,wang2020neural}, have been proposed. The RRT-based algorithms efficiently build a tree structure connecting the start and goal configurations by randomly sampling points within the configuration space. While sampling-based methods excel in handling high-dimensional or large-scale planning problems, their inherent randomization prevents them from finding the optimal path within a specific time constraint. 

After finding a collision-free path in the configuration space, trajectory optimization is applied to generate a safe, smooth, and dynamically feasible path for the vehicle~\cite{betts1998survey,toussaint2009robot}. It achieves this by formulating the path-smooth problem into constrained optimization problems. The objective is to optimize the smoothness of the path while incorporating various constraints to guarantee obstacle avoidance and adherence to vehicle dynamics. 
The complexity of these constraints determines the type of optimization problem used. Trajectory optimization can be formulated as non-linear programming~\cite{pardo2016evaluating}, mixed-integer programming~\cite{da2019collision}, or quadratic programming (QP) problems~\cite{he2021tdr,fan2018baidu}. QP offers the most efficient computation and is well-suited for highway and urban driving scenarios. These environments often have stricter time constraints but also benefit from structured layouts and predictable constraints.
%\vspace{-2em}

\subsection{QP Solvers for Path Planning}
This section dives into QP solving algorithms and software. We'll explore different QP solvers, examining their core techniques, strengths and weaknesses, and the key features needed for path optimization.

Many solvers exist to tackle QP problems. The qpOASES~\cite{qpOASES} solver utilizes the active-set method, which is a well-established method, working well for various problems. However, it can slow down significantly with large-scale scenarios, and behave badly for ill-conditioned problems or poorly chosen initial points. The OOQP~\cite{OOQP} solver utilizes interior point method, which iteratively solve linear equations obtained by a Newton-like method. The versatile solver incorporates techniques to deal with QP problems with various structures, such as sparse QPs and bound-constrained QPs. However, its reliance on repeatedly solving linear equations becomes a bottleneck for the massive problems encountered in autonomous driving. The OSQP~\cite{osqp} solver uses the alternating direction method of multipliers (ADMM)~\cite{ADMM}. By breaking down large problems into smaller, easier-to-solve pieces, OSQP can outperform other solvers tenfold in certain large-scale situations, while maintaining high accuracy. 

Autonomous driving demands real-time control and planning. Therefore, solver speed is a critical factor. OSQP's speed advantage with large-scale problems makes it the preferred choice for autonomous driving path planning~\cite{fan2018baidu}. 

\subsection{Path Planning System}
Commercial autonomous driving software, like Baidu Apollo~\cite{zhang2020apollo}, relies entirely on software for path finding and trajectory optimization. While this software-centric approach offers flexibility, it becomes increasingly difficult to manage real-time constraints as the complexity of autonomous driving systems grows~\cite{yu2020building}. Field-Programmable Gate Arrays (FPGAs) have emerged as promising hardware accelerators for various autonomous driving workloads~\cite{fang2017fpga,qin2019pi,wan2021energy,liu2021archytas,gan2021eudoxus,hao2022factor,liu2022energy,hao2023blitzcrank,hao2024orianna}. They offer significant performance benefits compared to software-only solutions. Previous research explored using FPGAs to accelerate Quadratic Programming (QP) solvers~\cite{jerez2011qpfpga}. These designs employed either interior point methods or active-set methods. 
%However, the scalability limitations associated with these methods make them unsuitable for the large-scale and real-time demands of autonomous driving.
However, existing work lacks exploiting problem-specific sparsity patterns and optimizing data flow at the system level, limiting its adaptation for large-scale and real-time problem solving on resource-constrained embedded platforms.

\section{Path Planning on FPGA: A Motivating Example} \label{sec:motivation}
End-to-end latency (from perception to action) is one of the most critical metrics for autonomous driving systems, as it has a significant impact on both safety and ride comfort. To quantify the latency requirement, we adopt the analytical model presented in \cite{yu2020building}:
\begin{align}
(T_{comp} + T_{data} + T_{mech}) \times v + \frac{1}{2} \times a \times T_{stop}^2 \leq D 
\\
T_{stop} = \frac{v}{a} 
\end{align}
Here, $T_{comp}$ denotes the time required for the computing system to process sensor inputs and generate control commands.
$T_{data}$ represents the time needed to transmit these commands to the vehicle's actuators via the vehicle Controller Area Network (CAN) bus.
$T_{mech}$ is the time for the mechanical components of the vehicle to start reacting. For a vehicle traveling at 36 km/h, every additional 100 ms of $T_{comp}$ increases approximately one meter of reaction distance. $T_{comp}$ includes perception, planning, and control, so the ideal planning time should be <100ms.

Traditional computing platforms, primarily general-purpose CPUs and GPUs, often struggle to meet the stringent real-time performance and power efficiency requirements demanded by sophisticated path planning algorithms deployed on embedded autonomous systems. While GPUs offer significant parallelism, they can consume substantial power, which is a critical constraint for battery-operated mobile platforms. Furthermore, the latency characteristics of GPU execution might not always align with the tight deadlines of real-time control loops in robotics. Path planning algorithms can require computation times ranging from hundreds of milliseconds to seconds on CPUs or GPUs, potentially hindering real-time responsiveness. Field-Programmable Gate Arrays (FPGAs) are emerging as a compelling alternative computing substrate for demanding robotics applications~\cite{RoboticonFPGAs}. FPGAs offer a unique combination of advantages well-suited to the challenges of real-time robotic computing: Energy Efficiency, and Hardware Customization.
% \cmt{Table~\ref{tab:other_method} compares the path planning
% performance with recent solutions on embedded FPGA. The results show that our optimization-based approach can generate collision-free, dynamically feasible, and reasonably smooth and efficient trajectories with real-time speed.}

Table~\ref{tab:other_method} compares the path planning performance with recent solutions on embedded FPGA. Our optimization-based approach explicitly incorporates dynamic feasibility and collision checking based on vehicle curvature limits as constraints in the problem formulation. Our method can generate collision-free, dynamically feasible, and reasonably smooth trajectories in real-time, while other methods fail to meet the above requirements.
\begin{table}[]
\centering
\caption{Path Planning Performance Comparison with other Path Planning Approches on embedded FPGA}
\label{tab:other_method}
{%
\begin{tabular}{|c|c|cc|c|}
\hline
              & \textbf{This Work} & \multicolumn{2}{c|}{\textbf{TC'24 \cite{p3netTC}}}       & \textbf{VLSIC'20 \cite{Astar_fpga}} \\ \hline
Method        & Optimization-based & \multicolumn{1}{c|}{Neural-based} & Sampling-based & Search-based              \\ \hline
Algorithm         & ADMM               & \multicolumn{1}{c|}{P3Net}        & BIT*~\cite{bitstar}           & Customized A*               \\ \hline
Platform &
  \begin{tabular}[c]{@{}c@{}}AMD ZCU102\\ FPGA @250MHz\end{tabular} &
  \multicolumn{1}{c|}{\begin{tabular}[c]{@{}c@{}}AMD ZCU104\\ FPGA @200MHz\end{tabular}} &
  \begin{tabular}[c]{@{}c@{}}Intel XeonW-2235\\ CPU @3.8GHz\end{tabular} &
  \begin{tabular}[c]{@{}c@{}}AMD ZCU102\\ FPGA @200MHz\end{tabular} \\ \hline
Planning Size & 700x700            & \multicolumn{1}{c|}{40x40}        & 40x40          & 1200x1200                 \\ \hline
Planning Time & \textbf{17ms }              & \multicolumn{1}{c|}{62ms}         & 1.08s          & 503ms                     \\ \hline
Kinetic Feasibility      & \Checkmark                & \multicolumn{1}{c|}{\XSolidBrush}           & \XSolidBrush             & \XSolidBrush                        \\ \hline
Collision-free   & \Checkmark                & \multicolumn{1}{c|}{\XSolidBrush}           & \XSolidBrush             & \XSolidBrush                        \\ \hline
\end{tabular}%
}
\vspace{-1.5em}
\end{table}

\section{Path Planning Algorithm Design} \label{sec:framework}\

This section first introduces the software pipeline of our path planning algorithm. We will then explore the formulation of the QP problem that smooths the planned trajectory. Subsequently, we'll delve into the details of the QP solver, providing a solid algorithmic foundation for our hardware implementation.

\subsection{Software Pipeline}
\label{sec:software_pipeline}
Our path planning system leverages real-time data from upstream modules in autonomous driving systems, including:
1) obstacle position and shape data, 2) ego vehicle position data, 3) the goal position and the way points and 4) map data. It generates smooth and collision-free paths. To achieve this, we propose a two-stage path planning system based on search and QP techniques. The algorithm pipeline is illustrated in Figure~\ref{fig:framework} and consists of the following three main steps.

\begin{enumerate}
    \item \emph{B-Spline Curve Generation}: 
    The goal of this step is to establish a baseline path that simplifies the subsequent searching and optimization process by limiting the search space. To achieve this, we account for the map and way-point information, and leverage B-spline curves to fit the way points. B-spline curves are ideal for robotics and autonomous driving path planning due to their computational efficiency and convex hull property~\cite{maekawa2010curvature}. 
   
    \item \emph{Dynamic Programming (DP) Search}: This step refines the initial path generated in step 1 to ensure both efficiency and collision-free. We incorporate obstacle data to identify areas to avoid. We discretize the driving space around the B-spline curve, creating a grid representation of the environment. We use Dijkstra's algorithm to search for the shortest collision-free path in this discretized space. Finally, we use a cubic spline curve to smooth the path.
    %The B-spline curve from Step 1 serves as a guide for the search, ensuring the resulting path remains close to the original trajectory while avoiding obstacles.
    \item \emph{Reference Path Processing}:
    This step processes the generated path in step 2, including adjusting the path point number and interval according to the setting. We also update more detailed obstacle information for the actual vehicle (front and rear). Then we generate all problem matrices for the final step.
    \item \emph{QP Optimization}: The final step polishes the path obtained in step 3, guaranteeing it's not only collision-free but also dynamically feasible for the vehicle to follow. To achieve this goal, we formulate this step as a constrained QP optimization problem and use ADMM~\cite{osqp} based algorithm to solve it. 
\end{enumerate}

\subsection{QP Problem Formulation}
\label{sec:qp}

Convex Quadratic Programming(QP) problems with $n$ decision variables and $m$ constraints are defined as follows:
\begin{align}
\text{Minimize} \quad & \frac{1}{2} \mathbf{x}^T \mathbf{P} \mathbf{x} + \mathbf{q}^T \mathbf{x}
\label{eq:qp_obj}
\\ 
\text{Subject to} \quad &  \mathbf{l} \leq \mathbf{Ax} \leq \mathbf{u} \label{eq:qp_cons}
\end{align}

\begin{figure}[t]
    \centering
    \includegraphics[width=\linewidth]{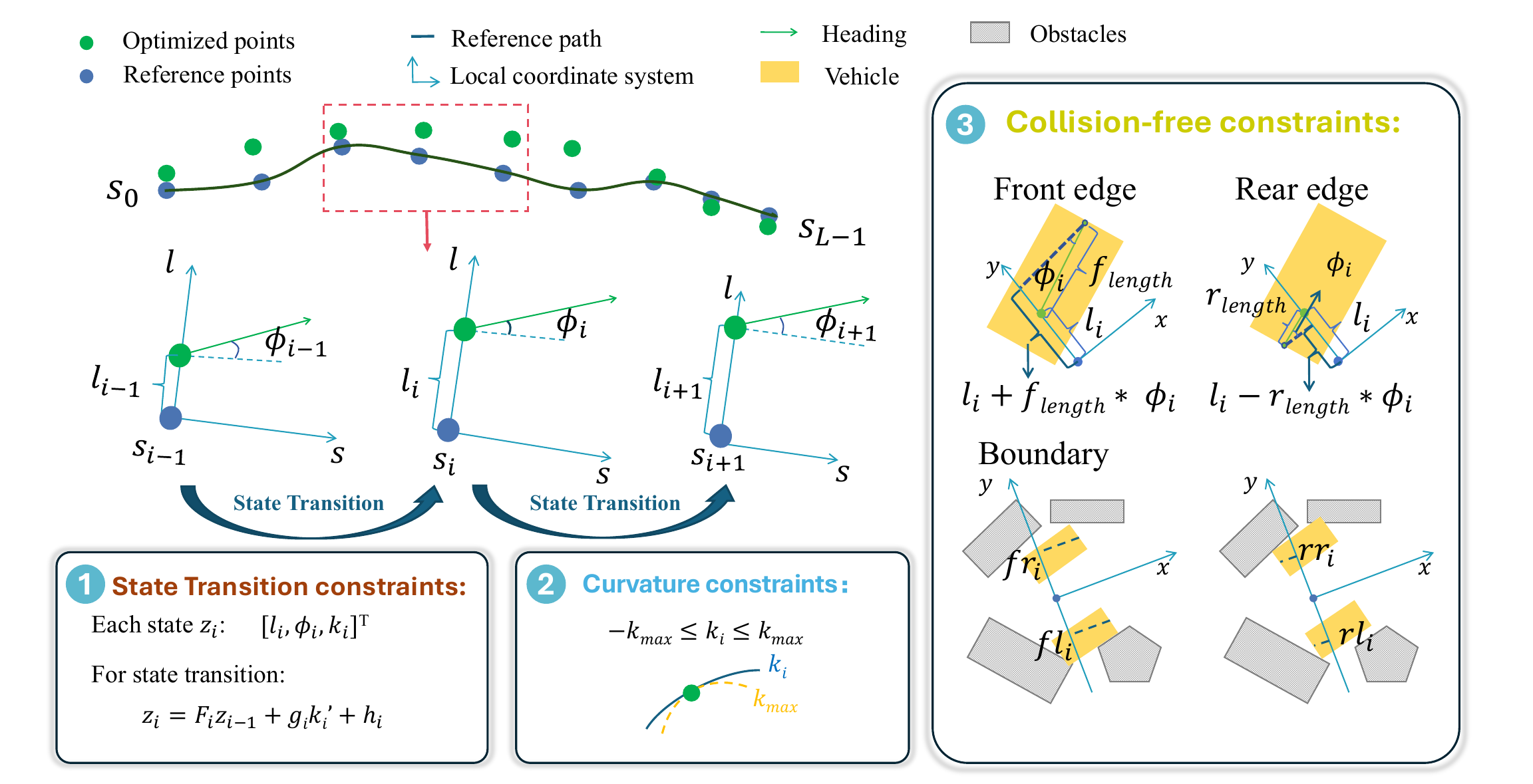}
    %\caption{System Architecture of Aplus}
    \caption{Trajectory Optimization and Its Constraints.}
    \Description{This figure shows the constraints in the path planning optimization problem. }
    \label{fig:cons}
\end{figure}

In the cost function Eq.~\ref{eq:qp_obj}, $x \in \mathbb{R}^{n}$ is the vector of decision variables (i.e., problem solution), where the positive semi-definite $P\in \mathbb{S}^{n}_{+}$ matrix and vector $q \in \mathbb{R}^{n}$ define the QP objective. 
In Eq.~\ref{eq:qp_cons}, the matrix $A\in \mathbb{R}^{m\times n}$ and vectors $\{l,u \}\in \mathbb{R}^{m}$ describe the problem constraints. 
% where $x \in \mathbb{R}^{n}$ is the vector of decision variables, the positive semi-definite $P\in \mathbb{S}^{n}_{+}$ matrix and vector $q \in \mathbb{R}^{n}$ define the objective, and the matrix $A\in \mathbb{R}^{m\times n}$ and vectors $\{l,u \}\in \mathbb{R}^{m}$ describe the constraints. 
We formulate the trajectory optimization problem into a QP problem. We extract sample points from the reference path generated by the DP process. At each sample point, we establish a local coordinate frame. The origin of this frame aligns with the vehicle's rear axle center. The x-axis aligns with the tangent of the reference path at that point, and the y-axis aligns with the normal direction. This local frame is illustrated in Figure~\ref{fig:cons}. For each sample point, we define its state as a vector $z_i=[l_i, \phi_i, k_i]^T$. $l_i$ represents the distance the optimized point can move along the y-axis relative to the reference point. $\phi_i$ represents the angle between the vehicle's heading and the x-axis of the local frame. $k_i$ represents the curvature of the optimized point. Our goal is to generate a path that is both smooth and collision-free. We achieve this by formulating an objective function as follows, which will be minimized during the optimization process,
\begin{equation}
Cost = w_l \sum_{i=0}^{L-1} l_i^2 + w_k \sum_{i=0}^{L-1} k_i^2 \\
+ w_{dk} \sum_{i=1}^{L-1} k_i'^2 + w_s \sum_{i=0}^{L-1}  \\(\varepsilon_{i,1}^2 + \varepsilon_{i,2}^2)
\label{eq:cost_func}
\end{equation}
where $w_l$, $w_k$, $w_{dk}$ and $w_{s}$ are hyper-parameters, $k_{i}'$ is the derivative of curvature $k_{i}$, $L$ is the number of samples. The first term penalizes large deviations of the optimized points from the samples. The second optimizes the overall smoothness of the path. The third term optimizes the smoothness of the path's curvature. The forth term ($\varepsilon
_{i,1}$ and $\varepsilon
_{i,2}$) introduces additional offsets to the vehicle's heading and rear end. This allows for slight adjustments to avoid obstacles while relaxing the strict positional constraints. We can further formulate Eq.~\ref{eq:cost_func} into an explicit quadratic form as follows,
 \begin{align}
Cost=&\sum_{i=0}^{L-1}{z_i^T diag\{w_l, 0, w_k\}z_i} + 
    \sum_{i=1}^{L-1}{k_i'w_{dk}k_i'} +  \sum_{i=0}^{L-1}{[\varepsilon_{i,1},\varepsilon_{i,2}]diag\{w_s,w_s\}[\varepsilon_{i,1},\varepsilon_{i,2}]^T} \label{eq:cost_1}\\
    =& x^T diag\{w_l, 0, w_k, ..., w_{dk},..., w_s,w_s, ...\}x \\
    =&x^T P x
\end{align}
where $x=
{[z_0^T,...,z_{L-1}^T,k'_1,...,k'_{L-1},\varepsilon_{0,1},\varepsilon_{0,2},..., 
\varepsilon_{L-1,1},\varepsilon_{L-1,2}]}^T$ and $P$ matrix is a $6*L-1$ dimensional square matrix.

To ensure the optimized path adheres to the vehicle's physical limitations, we incorporate curvature constraints into the QP formulation for each sample point, as follows, 
\begin{align}
    \label{eq:k_cons}
    -k_\text{max} \leq k_i \leq k_\text{max}.
\end{align}
where $k_\text{max}$ is the maximum curvature. $k_\text{max} =tan(\alpha_{max})/d$, where $\alpha_{max} $ is the maximum steering angle, and $d$ is the distance between the front and rear wheel.

To guarantee a collision-free path, we incorporate obstacle constraints into the QP formulation. These constraints leverage the minimum distances between the vehicle and obstacles within the local frame. As shown in Figure~\ref{fig:cons}, we use $fl_i$ and $fr_i$ to denote y-axis coordinate boundaries at the front edge of the vehicle, and use $rl_i$ and $rr_i$ to denote y-axis coordinate boundaries at the rear edge of the vehicle. The obstacle collision constraints are formulated as follows,
\begin{align}
\label{eq:boundary_cons1}
fl_i \leq l_i + f_\text{length} * \phi_i + \varepsilon
_{i,1} \leq fr_i  \\
\label{eq:boundary_cons2}
rl_i \leq l_i - r_\text{length} * \phi_i + \varepsilon
_{i,2}\leq rr_i
\end{align}
where $f_\text{length}$ and $r_\text{length}$ are the distances between the vehicle's rear axle center to the front and rear of the vehicle. Because $\phi_i$ is small, we use $\phi_i$ to approximate $sin \phi_i$ in the above equation.

To ensure the optimized path adheres to the vehicle's physical limitations and maintains coherence with the reference path, we incorporate spatial dynamic constraints into the QP formulation. For the vehicle trajectory points with state $z_i=[l_i, \phi_i, k_i]^T$, we construct a discretized state transfer equation for two adjacent trajectory points as follows:
\begin{align}
    \label{eq:dynamic_cons}
    z_i=F_i z_{i-1}+g_i k_i'+h_i,
\end{align}
where $F_i$, $g_i$ are the state transition matrix and control input matrix, both derived from the state Jacobian matrix. $g_i k_i'$ means that the control input only affects the derivative of the curvature $k_i'$. $h_i$ is a constant term ensuring the continuity of the state equation.

% where $F_i$, $g_i$ and $h_i$ are obtained from DP samples, capturing the relationship between the current state $z_i$ and the previous state $z_{i-1}$.

All the constraints introduced from Eq.\ref{eq:k_cons} to Eq.~\ref{eq:dynamic_cons} can be expressed in a standard form suitable for QP problems, $l \leq Ax \leq u$. In our problem settings, the constraint matrix $A$ is with $6 * L+2$ rows and $6 * L-1$ columns. With the objective function and the constraints, the path optimization is essentially to find the optimal state vector $x$ that minimizes the objective function, adhering to all the defined constraints. This minimization process is efficiently handled by QP solvers.

\subsection{QP Solving Using ADMM and PCG}

To solve the formulated QP problem efficiently, we employ an ADMM-based QP solver. This solver leverages the Alternating Direction Method of Multipliers (ADMM) algorithm, which is well-suited for handling problems with complex constraints. Before applying the ADMM algorithm, we first utilize a preconditioning technique on the matrices involved ($P$ and $A$, as derived in Section~\ref{sec:qp}). Preconditioning essentially scales the elements of these matrices to enhance numerical stability during the optimization process. The ADMM algorithm relies on solving linear equations iteratively to reach the optimal solution. However, instead of using traditional matrix decomposition methods, we leverage Preconditioned Conjugate Gradients (PCG) for solving these linear equations within the ADMM framework.

\subsubsection{Preconditioning}
While the ADMM algorithm offers numerous advantages for solving QP problems, it's important to acknowledge a known limitation: its handling of "ill-conditioned" problems. These problems can cause the ADMM algorithm to converge slowly or even fail to converge entirely. To address this challenge, we introduce a preconditioning method that uses the matrix equilibration technique~\cite{Ruiz2001ASA} to accelerate the convergence of ADMM in our application.

%% \vspace{-0.9em}
\begin{algorithm}
\caption{ADMM Algorithm}
\label{alg:ADMM}
\begin{algorithmic}[1]

\STATE Given initial $x^0$, $z^0$, $y^0$, and parameters $\rho > 0$, $\sigma > 0$, $\alpha \in (0,2)$
\REPEAT
%\STATE $(\widetilde{x}^{k+1}, \widetilde{v}^{k+1}) \leftarrow$ solve 
\STATE solve \label{alg:line:linear}
$\left(P+\sigma I+\rho A^T A\right)\widetilde{x}^{k+1}=\sigma x^k-q+A^T\left(\rho z^k-y^k\right) $ \quad   $\blacktriangleright$ \textbf{Algorithm~\ref{alg:PCG}} 
\STATE $\widetilde{z}^{k+1} \leftarrow A \tilde{x}^{k+1}$
\STATE $x^{k+1} \leftarrow \alpha \widetilde{x}^{k+1}+(1-\alpha)x^k$
\STATE $z^{k+1} \leftarrow \Pi(\alpha \widetilde{z}^{k+1}+(1-\alpha)z^{k}+\rho ^{-1}y^{k}) $
\STATE $y^{k+1}=y^{k}+\rho(\alpha \widetilde{z}^{k+1}+(1-\alpha)z^{k}-z^{k+1})$

\UNTIL{termination criterion is satisfied}
\end{algorithmic}
\end{algorithm}
%% \vspace{-0.83em}
\subsubsection{ADMM Alogrithm}
The ADMM algorithm used for solving the QP path optimization is described in Algorithm~\ref{alg:ADMM}. 
In the algorithm, ADMM transforms QP problems into an iterative process of solving linear equations and vector updates. These operations are generally less computationally expensive compared to traditional QP solvers, making ADMM suitable for real-time applications on embedded systems with limited resources.

\subsubsection{PCG for Solving Linear Systems}
The ADMM algorithm, while powerful, relies on solving linear systems as a key step. However, solving these systems can become computationally expensive, especially for large-scale problems encountered in trajectory optimization. The computational cost of traditional methods like matrix decomposition approaches (e.g., LDL decomposition) becomes prohibitively large when the linear system scales~\cite{GPUAcc}.
To address this challenge, we employ the PCG method, detailed in Algorithm~\ref{alg:PCG}, for solving the linear systems within the ADMM algorithm. PCG offers several advantages for our application: 1) it is specifically designed for efficiently handling large, sparse linear systems, which are typical in trajectory optimization; 2) As the algorithm shows, the PCG is an iterative method that mainly involves matrix-vector multiplication, vector scalar multiplication (AXPY), and dot-product. These operations can be effectively parallelized with the massive parallelization capabilities of FPGAs.

\begin{algorithm}
\caption{Preconditioned Conjugate Gradients (PCG) Method}
\label{alg:PCG}
\begin{algorithmic}[1]
\STATE Linear system $Kx = b$, \\
Jacobi preconditioner $M=diag(K_{00}, K_{11},...)$ 
\STATE Initial $x ^ 0 = 0, r^0 = b-Kx^0 , y^0=M^{-1}r^0,p^0=y^0,k=0$
\WHILE{$||r^k||>\epsilon ||b||$}
\STATE $\alpha^k \leftarrow \frac{((r^k)^T)y^k}{(p^k)^TKp^k}$
\STATE $x^{k+1} \leftarrow x^{k}+ \alpha^{k}p^k$ 
\STATE $r^{k+1} \leftarrow r^{k} -\alpha^{k}Kp^k$ \label{step:r_pcg}
\STATE $y^{k+1} \leftarrow M^{-1}r^{k+1}$\label{step:y_pcg}
\STATE $\beta^{k+1} \leftarrow \frac{((r^{k+1})^T)y^{k+1}}{(r^k)^Ty^k}$
\STATE $p^{k+1} \leftarrow y^{k+1}+ \beta^{k+1}p^k$
\STATE $k\leftarrow k+1 $
\ENDWHILE
\end{algorithmic}
\end{algorithm}

%To enhance the convergence speed of ADMM, it's crucial to set a sufficiently large $\rho$ for equation constraints, yet an overly large $\rho$ could hinder PCG convergence. Therefore, we have made improvements to the setting of $\rho$. The improved parameter is:
%\begin{align}
%\rho=diag(\rho_1,...,\rho_m), \quad \rho_i=
%    \begin{cases}
%		\bar{\rho} \quad  & l_i \neq u_i \\
%		10\bar{\rho} \quad  &  l_i = u_i 
%    \end{cases} \text{,}
%    \label{eq:rho_i}
%\end{align}
%where $\bar{\rho}>0$ is initialized to 0.1. This approach guarantees the convergence speed of PCG. However, it may result in a slowdown of ADMM convergence due to the initialization scheme. Thus, in this design, $\bar{\rho}$ is updated every 10 iterations to maintain the convergence speed of ADMM. This initialization and update strategy ensures simultaneous convergence speed for both ADMM and PCG. 

\section{Path Planner Architecture Design}\label{sec:aplus}

\begin{figure}
    \centering
    \includegraphics[width=0.9\linewidth]{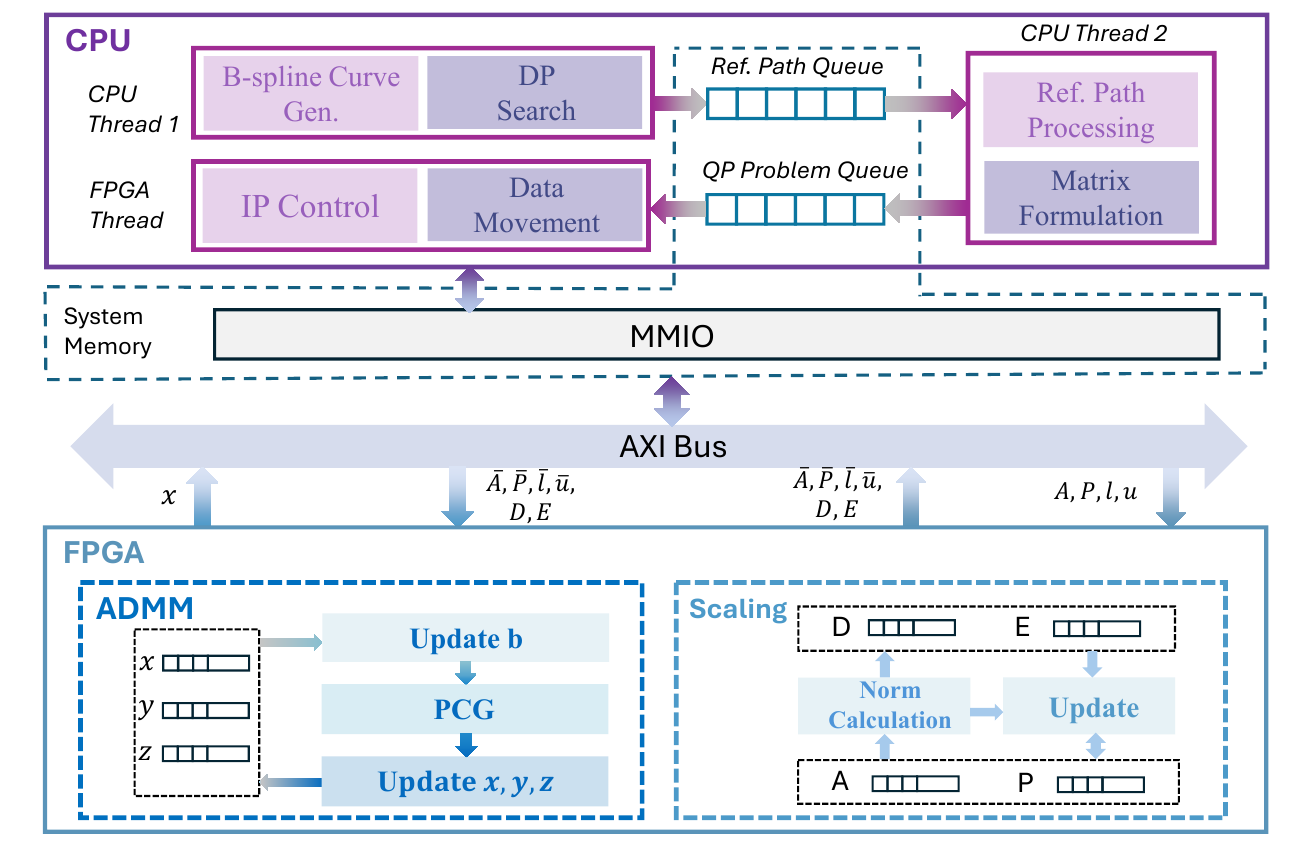}
    %\caption{System Architecture of Aplus}
    \vspace{-1em}
    \caption{System Architecture of Proposed Path Planner}
    \Description{This figure shows the system architecture of the proposed path planner. }
    \label{fig:hardware_frame}
\end{figure}

The path planning system framework of this design first performs DP search to transform the problem into convex optimization, and then uses the QP optimization method to solve the path. In the DP process, the search interval along the spline curve direction will be relatively small, while the sampling points in the QP process will be dense, and the dimension of the QP problem will increase linearly. Therefore, the solving time of the QP problem will be a decisive factor in the solving time of the entire path planning solver. Therefore, this design will partially deploy the solution to the quadratic programming problem on FPGA, as shown in Figure~\ref{fig:hardware_frame}. 
The implementation of the QP process includes two parts: Scaling and ADMM. The Scaling module scales the data of the QP problem, writes the scaled data back to DDR, and the ADMM module reads the scaled matrices for solving. 

%To enhance performance and efficiency further, the design stores reference samples in DDR (Dual Data Rate) memory. This not only conserves valuable memory space but also boosts data access speed, facilitating rapid acquisition and processing of necessary reference data. This becomes particularly critical when handling substantial datasets, significantly reducing data transmission times and elevating the overall system responsiveness.

%Another pivotal aspect of this design is the hardware implementation of the computationally intensive quadratic programming (QP) process on the FPGA's programmable logic (PL) side. This step transforms the QP algorithm into hardware and harnesses the FPGA's parallel computing power, further enhancing system performance.

%In summary, this design optimally allocates key functions across diverse segments of both the CPU and FPGA, fully leveraging their unique traits and performance advantages. This results in the system efficiently generating B-spline curves, conducting DP searches, and executing QP processes, ultimately achieving notable acceleration effects.
\subsection{Problem Matrix Formulation}
\label{sec:problem_matrix}
\begin{figure}[h!]
    \centering
\includegraphics[width=\linewidth]{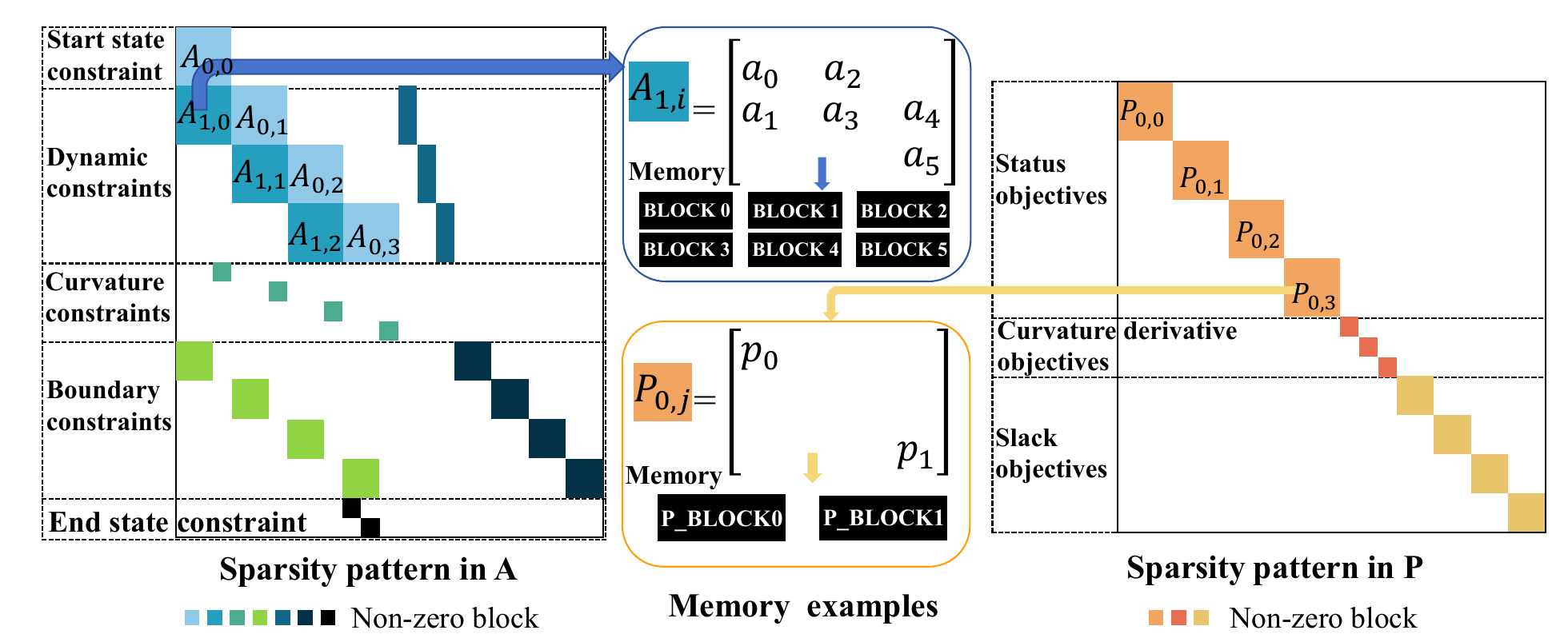} 
    \caption{The sparsity pattern of matrix $A$ and $P$}
    \label{fig:AP_memory}
\end{figure}

The problem matrices $A$, $P$ are involved throughout the algorithm. Therefore, we analyze the problem-specific information in $A$ and $P$ and look into the customization opportunity. Figure~\ref{fig:AP_memory} shows the non-zero elements distribution in $A$ and $P$, when the number of reference points is 4. As discussed in section~\ref{sec:qp}, the cost matrix $P$ is derived from eq.~\ref{eq:cost_1}, and the constraint matrix $A$ from eq.~\ref{eq:k_cons} to eq.~\ref{eq:dynamic_cons}. We observe a structural sparse pattern in the problem matrices, and the matrices scale with the number of trajectory points. Therefore, we propose a sparse pattern-aware storage scheme, as shown in Figure~\ref{fig:AP_memory}. $A_{1, j}(j=0,1,2)$ are all 3$\times$3 blocks with the same pattern. The non-zero elements at position $a_1$ in $A_{1, j}(j=0,1,2)$ are stored in A\_Block\_0, which means using 6 memory blocks to store all non-zero elements in $A_{1, j}(j=0,1,2)$. We store all non-zero blocks in the same way. Finally, the matrix $A$ will use 17 memory blocks, matrix $P$ will use 5 memory blocks, and the size of each memory block will be the number of trajectory points $L$.
\subsection{Design of the Scaling Module}

The scaling module scales matrices $A$ and $P$ based on the infinite norm of the column vectors, avoiding excessive values to improve the convergence of the ADMM algorithm. There are two important steps in this module: Calculating the column norm of $A$, $P$, $A^T$ matrices to get the diagonal matrix for scaling; left/right-multiplying the diagonal matrix with $A$, $P$ to perform scaling. Because we use a fully decoupled matrix storage scheme shown in Figure~\ref{fig:AP_memory} (i.e., the elements in each row/column are stored in different memory blocks without dependencies), we can easily access a column/row of the matrix simultaneously, to compute the infinity norm or left/right-multiplication with diagonal matrices.
% By simultaneously traversing the non-zero elements in matrices $A$ and $P$, six diagonal elements in matrices $D$ and $E$ can be calculated for each cycle, and the results of the same cycle should be stored in different memory blocks. 
% \begin{figure}
%     \centering
%     \subfigure[The sparsity pattern of matrix $A$ and $P$.]{\includegraphics[width=\linewidth]{Figures/TACO/A_Pmemory_new_crop.pdf} \label{fig:AP_memory}}
%     \vspace{-1em}
%     \newline
%     \subfigure[Hardware architecture of of the Scaling module.]{\includegraphics[width=\linewidth]{Figures/TACO/scale_v2_crop.pdf}\label{fig:scale}}
%     \vspace{-1.5em}
%     \caption{Design of the Scaling Module}
%     \Description{This figure shows the Design of the Scaling Module. }
%     \label{fig:scale-design}
% \end{figure}
\begin{figure}[h]
    \centering
\includegraphics[width=\linewidth]{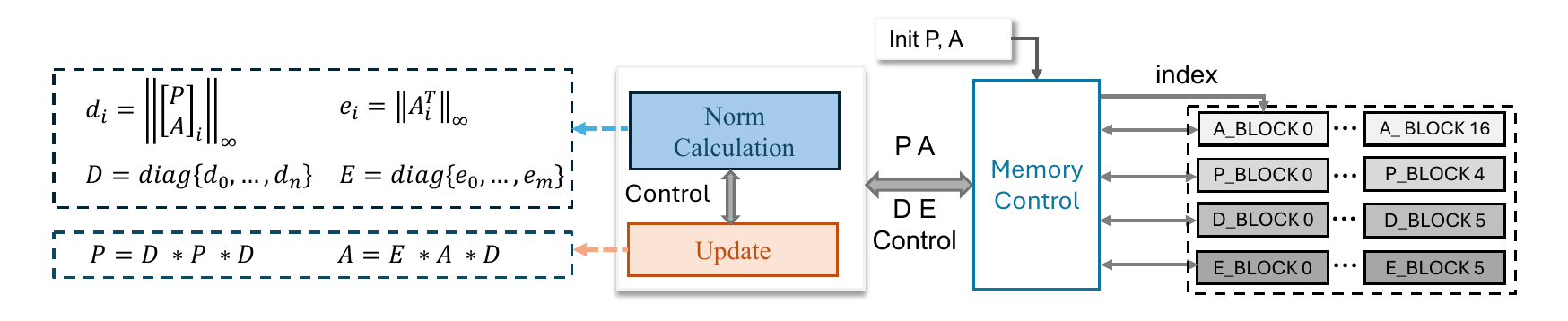}
    \caption{Hardware architecture of the Scaling module}
    \label{fig:scale} 
\end{figure}

The architecture of the scaling module is shown in Figure~\ref{fig:scale}. 
% In the norm calculation step, if matrices $D$ and $E$ need to be calculated simultaneously, it is necessary to customize the storage of matrix $P$ using the above methods, too. 
By simultaneously traversing the non-zero elements in matrices $A$ and $P$, in each iteration, we can calculate six elements of diagonal matrices $D$ and $E$ in parallel. The results should also be stored in different memory blocks, similar to $A$ and $P$.
In the update matrix step, we send the non-zero elements in matrix $D$ and $E$ to the update $P$ and update $A$ steps synchronously, achieving simultaneous updates of $P$ and $A$.

%The scaling module designed based on this storage method is shown in Figure~\ref{fig:scale}. When calculating the column vector norm of A, six results will be generated simultaneously. The storage method of the corresponding diagonal elements of the diagonal matrix and matrix P should also correspond to it. This design can simultaneously access all non-zero elements of multiple rows and columns under regular memory access conditions, ensuring alignment between computing and storage units, thus achieving efficient parallelization of computation. 

\subsection{Design of the ADMM}

\begin{figure}
    \centering
    \includegraphics[width=0.9\linewidth]{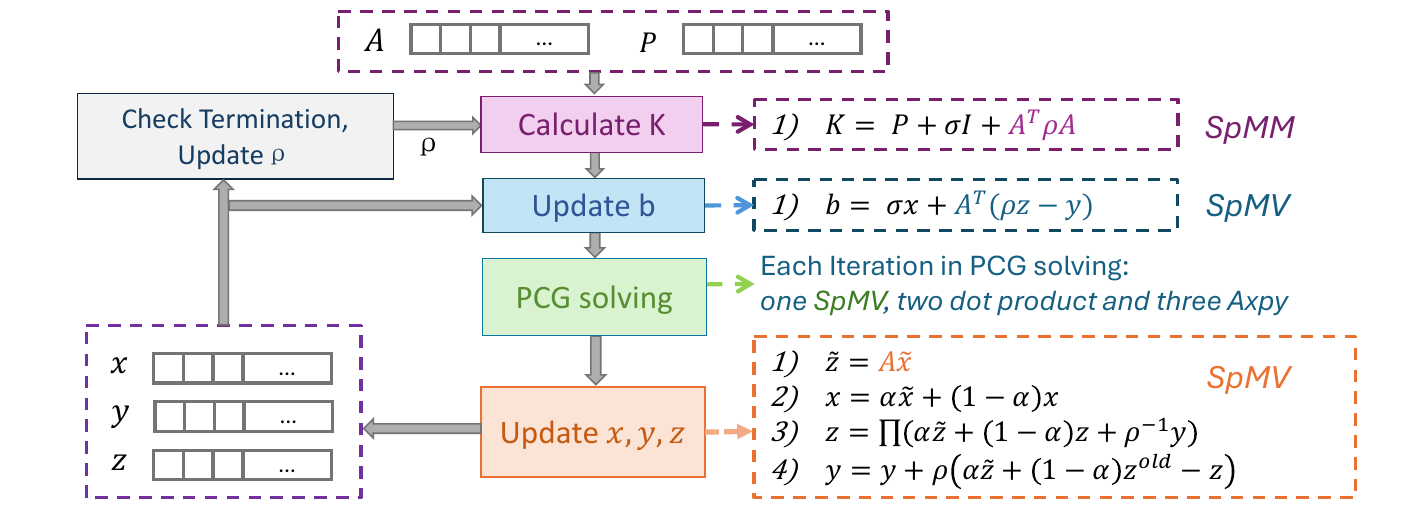}
    \caption{Architecture of the ADMM Module}
    \Description{This figure shows the Architecture of the ADMM Module. }
    \label{fig:ADMM}
    \vspace{-1em}
\end{figure}

The ADMM module can be divided into five parts: coefficient matrix calculation, $b$ update, PCG, vectors update, and termination check, as shown in Figure~\ref{fig:ADMM}. 
The coefficient matrix calculation part calculates the coefficient matrix $K$ based on matrices $A$, $P$ and $\rho$, and stores it in K\_MEM\_BLOCK. This part only needs to be called when updating $\rho$, as matrices $P$ and $A$ will not change during the iteration process. The $b$ update part calculates the vector $b$ based on $x, y, z$ in each iteration and stores it in b\_MEM\_BLOCK. The PCG module extracts $K$ and $b$ from K\_MEM\_BLOCK and b\_MEM\_BLOCK to solve the linear system $K \tilde{x}=b$. The vectors update part updates $x, y, z$ based on $\tilde{x}$ and $A$. The check part calculates residuals and updated $\rho$ based on matrices $A$, $P$ and vectors $x$, $y$, $z$. 

\subsection{Implementation of Preconditioned Conjugate Gradient (PCG) Algorithm}
\label{sec:PCG}

% \begin{figure}[ht]
%     \centering
%     \includegraphics[width=0.6\linewidth]{Figures/PCG_v2.pdf}
%     \vspace{-1em}
%     \caption{Block Diagram of PCG Method}
%     \Description{This figure shows the Block Diagram of PCG Method. }
%     \label{fig:PCG_diagram}
%     \vspace{-1em}
% \end{figure}
In ADMM Algorithm, solving the linear system (step~\ref{alg:line:linear}) occupies most of the computational workload, since other steps (updating $x,y,z$) only involve one sparse matrix-vector multiplication (SpMV) and three vector operations, while PCG solving requires n SpMV, 2n dot products, and 4n vector operations (n = \#iterations in PCG ). The above operations have more opportunities to be pipelined and parallelized.
% We can easily observe that the critical path of computation in the ADMM algorithm is the step~\ref{alg:line:linear}, since it has much greater time complexity than the remaining vector operations. Therefore, it is crucial to optimize the solution of linear systems, especially for large-scale, sparse systems. 
% In this paper, we use Preconditioned Conjugate Gradient (PCG) method. Algorithm~\ref{alg:PCG} shows how PCG works. The PCG algorithm mainly involves vector operations and matrix-vector multiplication, which have more opportunities to be pipelined and parallelized.
% Each PCG iteration includes one Matrix-Vector Multiplication, two Dot-Product Operations, and four Vector Operations. 
Additionally, we observe a specific sparse pattern in matrix K to be multiplied, as illustrated in Figure \ref{fig:ATA}. Based on these observations, we propose three optimizations for PCG solving: (1) Pipelined and parallelized processing units for vector linear operations and dot-product; (2) Pattern-aware Specialized Sparse Matrix-vector Multiplication (SpMV) Unit; (3) Algorithm optimization for faster convergence. We will mainly cover (2) and (3) since they demonstrate more novelty.

\section{Multi-level Optimizations for Path Planning Pipeline}
\label{sec:path_planner_opt}
Section~\ref{sec:aplus} gives a brief introduction of the proposed path planning framework. This section will illustrate the novel multi-level optimizations for path planning pipeline leveraging task and platform-specific information, including efficient sparse matrix storage and computing units, multi-level dataflow optimization, and design space exploration for algorithm-architecture co-optimizations.

\subsection{Sparsity-aware Hardware Design for Matrix Operations}
\label{sec:sparse_opt}
\subsubsection{Sparse Matrix Multiplication Optimization in ADMM}
\label{sec:MM}
\begin{figure}[ht]
    %\vspace{-1em}
    \centering
    \includegraphics[width=\linewidth]{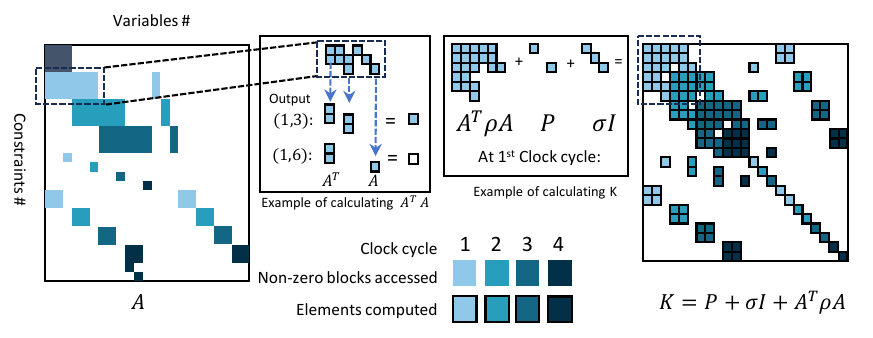}
    \caption{The process of calculating the coefficient matrix K.}
    \Description{This figure shows the process of calculating the coefficient matrix K.}
    \label{fig:ATA}
\end{figure}
To compute the coefficient matrix in ADMM, we must perform a matrix multiplication operation involving $A^T \rho A $. This calculation entails pairwise dot products between the columns of matrix $A$ and the columns weighted by the diagonal matrix $\rho$. We have devised an access scheme that extracts six columns of elements from matrix $A$ at a time, as shown in Figure~\ref{fig:AP_memory}. These six columns will produce a non-zero product with each other, but will not produce a non-zero product with other columns. Figure~\ref{fig:ATA} illustrates the computation process of $K$ when there are four reference points. The process is divided into four cycles, explaining the non-zero elements accessible in the matrix $A$ for each clock cycle and the corresponding non-zero elements that can be computed in the matrix $K$. This access plan takes into account the distribution characteristics of matrix $A$ to optimize access and computational efficiency. Firstly, during the matrix multiplication process, it is only necessary to traverse the non-zero elements of matrix $A$ once, demonstrating efficient access efficiency. Secondly, the non-zero elements of six columns in the $K$ matrix can be computed for each clock cycle. If the task latency of the $K$ computation process is $Latency1$ clock cycles, then the total latency of computing $K$ is $(L+Latency1-1)$ clock cycles. The number of non-zero elements in matrix $K$ is $36L-17$, indicating that the process of calculating $K$ achieves a high computational efficiency. 

\subsubsection{Efficient Sparse matrix-vector multiplication (SpMV)}
\label{sec:spmv}

\begin{figure}[ht]
    \centering
    \centering
    \includegraphics[width=\linewidth]{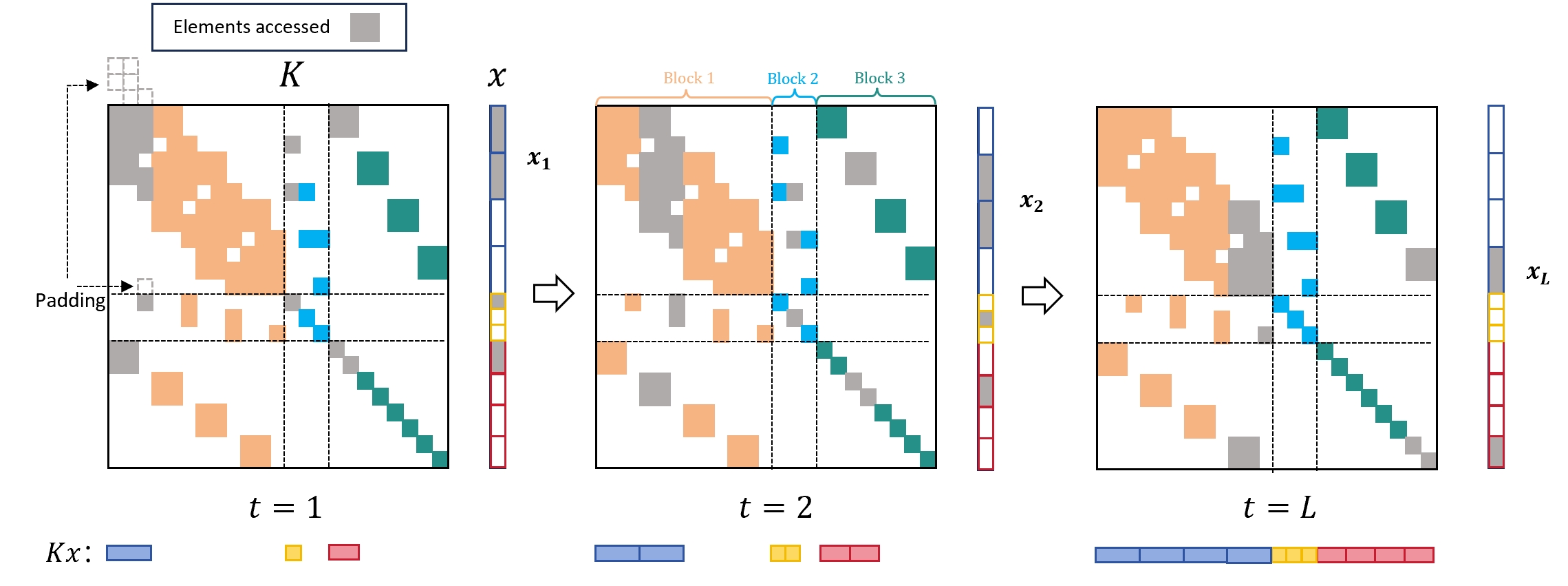}
    %\caption{The Structured Data Access for SpMV}
    \caption{Demonstration of proposed SpMV design}
    \Description{This figure shows the Structured Data Access for SpMV. }
    \label{fig:spmv_matrix}
    \vspace{-1em}
\end{figure}
In the ADMM algorithm, the matrix-vector multiplications are performed more frequently than matrix multiplication. Given the large matrix size (> $10^3$) and high sparsity (Non-Zero Elements < 0.5\% ), we proposed an efficient pattern-aware SpMV that enables high parallelism and structured memory access. First, we analyze the characteristics of the sparse matrix $K$. As discussed in section~\ref{sec:MM}, $K$ matrix is derived from $P$ and $A^T \rho A$, thus symmetric. Then we analyze $K$'s sparsity pattern and design a specialized hardware accordingly. As shown in Figure~\ref{fig:spmv_matrix}, we use a simplified
version of $K$ to demonstrate how our proposed pattern-aware SpMV works.
We divide $K$ into three blocks column-wise, each with a specific sparsity pattern. For block 1, we observe that every three columns have the same number of non-zero elements. Since we use the Compressed Sparse Column (CSC) format for sparse matrix $K$, we can obtain three columns by continuously fetching a fixed number of elements. In this way, we can access elements in CSC format without using row index and column pointers, which introduces an uncertain loop bound and is very inefficient for FPGA. Besides, since $K$ is symmetric, we can use a column in $K$ multiplied by $x$ to get an output. We further partition $K$ and vector $x$, so that every iteration is fully parallelized. We also store the partial sum for the next iteration to save hardware resources. Block 2 and Block 3 follow the same fashion, and we calculate 1 and 2 columns respectively. Therefore, for each clock cycle, we can get 6 (3+1+2) output vector elements. In general, we significantly improve the performance of SpMV and achieve high utilization of computing units since we fully pipeline and parallelize the whole computation.

\subsection{Multi-Level Dataflow Optimization and Operator Fusion}
\label{sec:dataflow_opt}
Section~\ref{sec:sparse_opt} proposes the problem-specific sparsity-aware hardware design. It focuses on optimizing individual operators (e.g., SpMV, SpMM). In this section, we will discuss higher-level optimizations, including inter-operator level dataflow optimization and system-level pipeline.

\begin{figure}[ht]
    \centering

    \includegraphics[width=\linewidth]{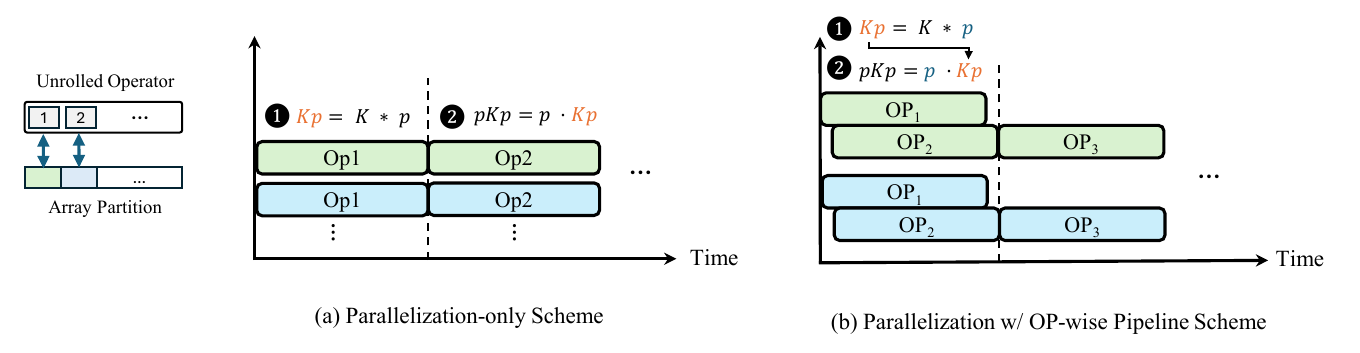}
    %\caption{The Structured Data Access for SpMV}
    \caption{An Example of fine-grained inter-operator pipeline (Operator Fusion) }
    \Description{ }
    \vspace{-1em}
    \label{fig:fused_operator}
\end{figure}
\subsubsection{Inter-Operator Level Dataflow Optimization}
The work~\cite{pathplanner_iccad} achieves inner-operator parallelization for vector and matrix operations that significantly improved performance. However, simply increasing the parallelism of individual operators will result in a proportional increase in resource usage, which could be marginally inefficient on embedded platforms. Additionally, in \cite{pathplanner_iccad}, all operators are executed sequentially at the algorithm level, leaving a huge space for fine-grained dataflow optimization to overlap latency.
To further accelerate the computation, we analyze the operators involved in the ADMM algorithm. We observed that some operators (e.g. step~\ref{step:r_pcg} and \ref{step:y_pcg} in algorithm~\ref{alg:PCG}) have read-after-write dependencies only at inter-operator level, i.e., each output element of the former operator can be directly sent as input to the latter operator. In this case, we can apply a fine-grained pipeline across those operators (Operator Fusion). Figure~\ref{fig:fused_operator} gives an example of the proposed dataflow optimization combining inter-operator pipelining and inner-operator parallelization. With this approach, we can overlap the latency of multiple operators with negligible resource overhead. We will introduce the detailed implementations in the following sections. 
% Previous design achieves inner-operator parallelization that significantly accelerates PCG solving. However, all inter-operator-level operations happen sequentially, leaving huge space for fine-grained dataflow analysis and optimization. We further propose a inter-operator level dataflow optimization, as shown in Fig~\ref{fig:fused_operator}.We observe that in some vector operations, the elements only have inter-operator dependency instead of inner-operator dependency, i.e after a element in the vector is calculated, it can be directly send to next operator for inter-operator pipeline. 
\subsubsection{Optimized PCG Solving with Operator Fusion}
\begin{figure}[t]
    \centering

    \includegraphics[width=\linewidth]{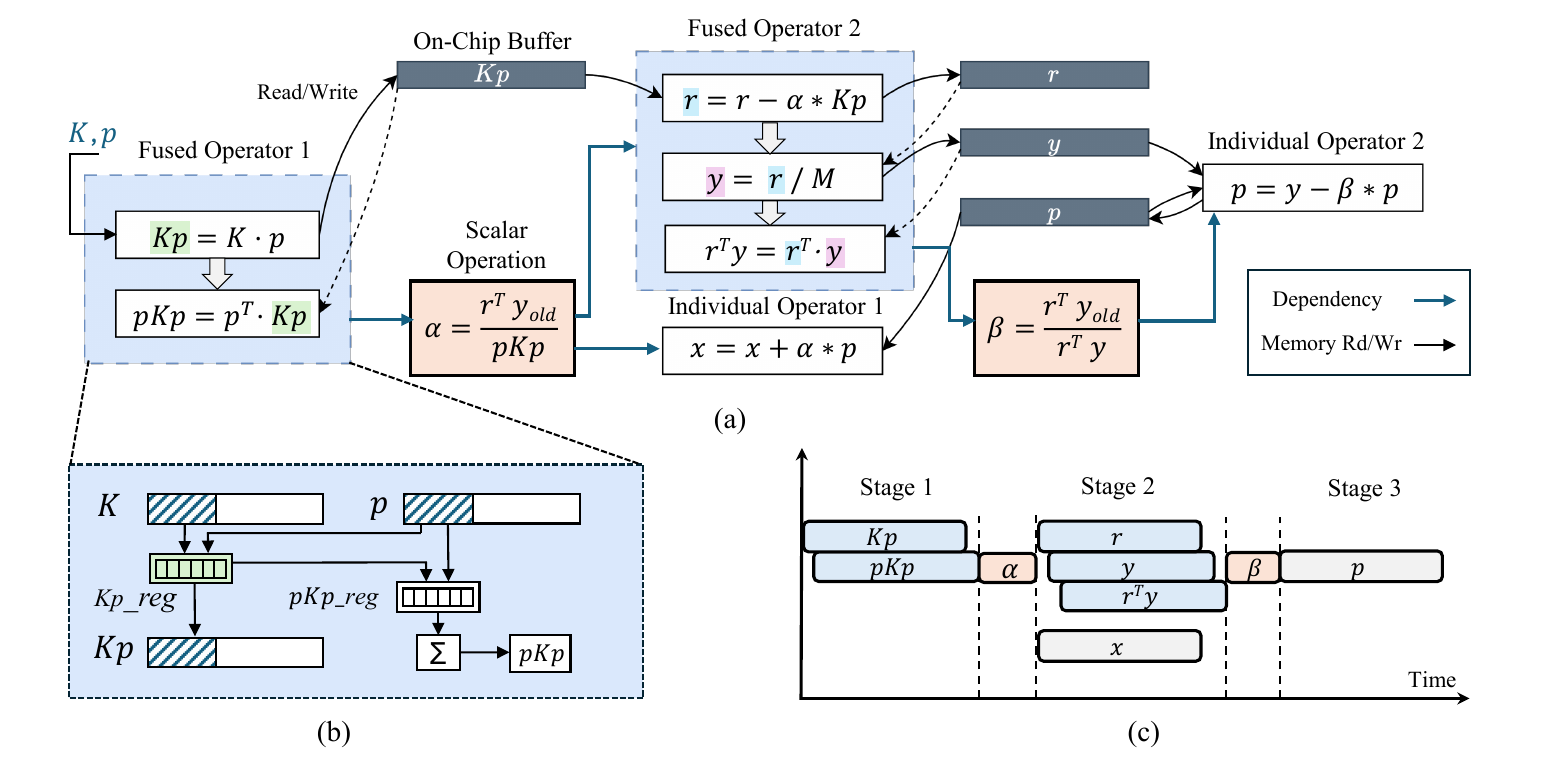}
    \vspace{-1em}
    %\caption{The Structured Data Access for SpMV}
    \caption{(a) Illustration of Operator Fusion in PCG algorithm (b) Details in fused operator (c) The overlapped computation flow}
    \Description{. }
    \vspace{-1em}
    \label{fig:pcg_op_fusion}
\end{figure}

% In this section, we apply the operator fusion technique in linear system solving (Algo~\ref{alg:PCG}). The algorithm~\ref{alg:PCG} contains 
Since we use the iterative method (PCG method, algorithm~\ref{alg:PCG}) to solve the linear systems in ADMM, overlapping the latency within one iteration is critical to accelerate the solution. One iteration in PCG contains seven major operators (one SpMV, two dot products, one element-wise multiplication, and three vector linear operations).   Figure~\ref{fig:pcg_op_fusion} illustrates how operator fusion works in the PCG algorithm. According to the computation graph, the two scalar operations (calculating $\alpha$ and $\beta$) divide the whole computation into three stages and act as the hard boundaries of pipelining. Inside each stage, we identify two groups of operators that can be fused, which are marked in a blue dotted box. Figure~\ref{fig:pcg_op_fusion} (b) shows the details of a fused operator. By adding registers to store the results of key variables (marked with color), we can directly pass those values to the next dependent operator in a clock cycle, without writing back and reading on-chip memory. Besides, since the parallelization factor is 6 or 12 for each operator, we only need 6/12 additional registers to implement the fine-grained pipeline. As Figure~\ref{fig:pcg_op_fusion} (c) illustrates, we can reduce up to 57\% latency per PCG iteration after performing operator fusion.

% \begin{figure}[ht]
%     \centering

%     \includegraphics[width=\linewidth]{Figures/hybird_dataflow_optimization_v3.png}
%     \vspace{-1em}
%     %\caption{The Structured Data Access for SpMV}
%     \caption{Illustration of Operator Fusion in PCG algorithm}
%     \Description{. }
%     \label{fig:pcg_op_fusion}
% \end{figure}

\subsubsection{Optimized Vectors Update in ADMM}
\label{sec:vec_update}
\begin{figure}[ht]
    \centering
    \includegraphics[width=\linewidth]{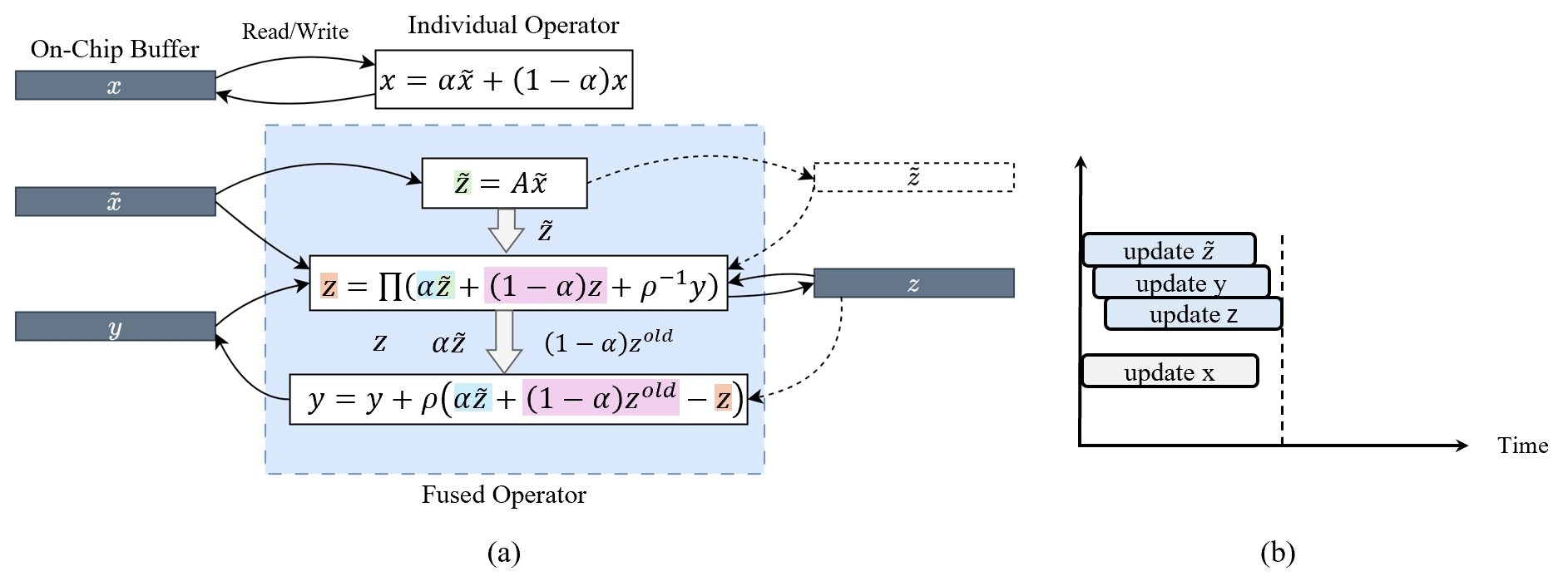}
    %\caption{The Structured Data Access for SpMV}
    \caption{(a) Illustration of Optimized Vector Update (b) The overlapped computation flow}
    \label{fig:vec_update_opt}
    \vspace{-1em}
    \Description{. }
\end{figure}

% After the PCG solving follows the vector update module has a similar computation pattern as vector operations in PCG. We also identify operator fusion opportunities to overlap latency.
In the vector update module after PCG, the naive implementation calculates $\tilde{z}$ and updates $x$, $y$, and $z$ sequentially. Through dependency analysis, we observe a computation pattern similar to the vector operations in PCG. Additionally, the intermediate values in \textit{update z} are also required in the \textit{update y}. Here we identify opportunities for operator fusion and data reuse. Figure~\ref{fig:vec_update_opt} demonstrates the optimization scheme for the vector update module. The reused variables and intermediate values are marked with colors. By introducing operator fusion, we reduce four vector memory read/write operations, on-chip memory array $\tilde{z}$, and two vector multiplications $\alpha \tilde{z}$ and $(1-\alpha)z$ with negligible overhead (four groups of registers), which significantly improves the computational efficiency on resource-constrained platform.

% In the vector update module, we first calculate $\tilde{z}$, and then sequentially update $x$, $y$, and $z$. During the update of $z$, it is necessary to use the old $y$ and $z$. Similarly, during the update of $y$, it is necessary to use the old $y$ as well as the old and updated $z$. In order to avoid multiple accesses to $y$ and $z$, we have implemented an optimization measure. Before writing the updated $z$ back into memory, it is fed into the update unit of $y$. In this process, we use registers to store the old $y$ and $z$. This strategy effectively reduces read operations on memory and improves computational efficiency. In Section~\ref{sec:optimization}, we compare the delay before and after optimization.

\subsubsection{System-level optimization for end-to-end throughput improvement}
\label{sec:systemlevel_pipeline}
\begin{figure}[ht]
    \centering
    \includegraphics[width=\linewidth]{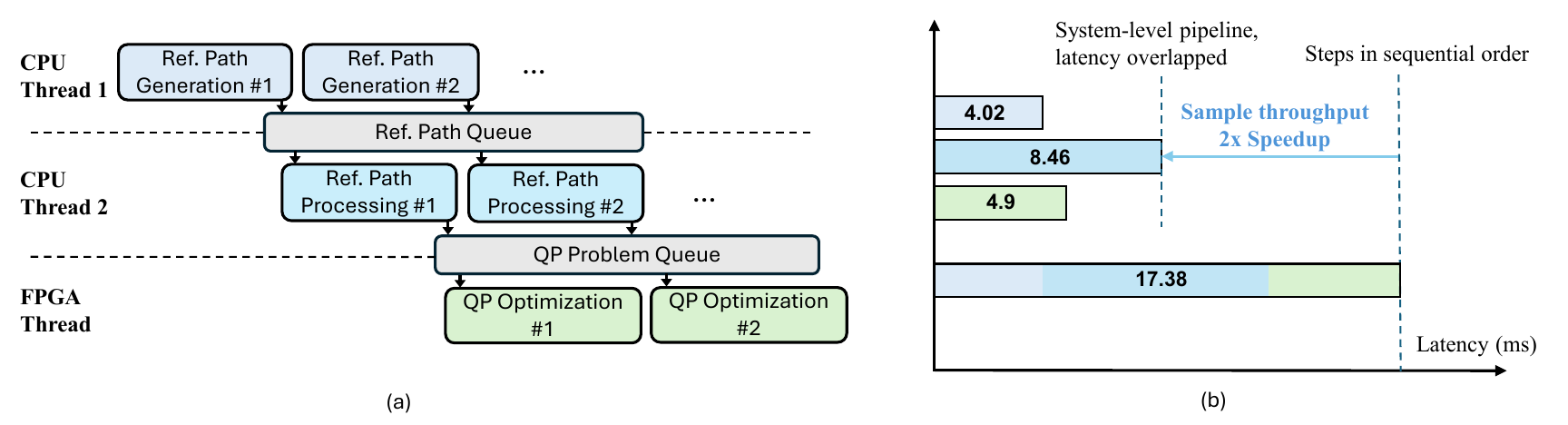}
    %\caption{The Structured Data Access for SpMV}
    \caption{System-level pipeline (a) multi-thread scheduling diagram (b) throughput improvement}
    \Description{. }
    \label{fig:system_level_pipeline}
\end{figure}

Previous work~\cite{pathplanner_iccad} focused on accelerating the path optimization module. However, as section~\ref{sec:software_pipeline} discusses, a typical path planning pipeline also includes reference path generation and pre-processing before path optimization. Failing to consider them system-wide will limit end-to-end throughput. 
We first profile the computation performance of all steps on CPU. Table~\ref{tab:software_breakdown} shows the execution time breakdown of each step. The Quadratic Programming takes up most of the execution time, indicating a dominant computing capacity requirement, while the reference path generation and pre-processing are less computationally intensive. Therefore, we use a multi-threaded heterogeneous scheme to improve system throughput. In the host program, we assign one thread for each module. The reference path generation and pre-processing modules are mapped on the CPU. We pipeline these modules using FIFO queues. Figure~\ref{fig:system_level_pipeline} shows our system-level pipeline. With all three modules overlapping, our design achieves 2$\times$ end-to-end throughput improvement. The communication overhead has been included in the result.

% \begin{table}[]
% \centering
% \caption{Computational workload of each module}
% \label{tab:workload}
% \resizebox{0.8\textwidth}{!}{%
% \begin{tabular}{ccccc}
% \hline
% Module                & \#Multiplication & \#Division & \#Addition & \#Comparison \\ \hline
% B-Spline Curve        & 510              & 255        & 561        & 0            \\
% Dijkstra              & 15147            & 15147      & 15147      & 10098        \\
% Quadratic Programming & 1386172          & 80         & 153380     & 234950       \\ \hline
% \end{tabular}%
% }
% \end{table}

\begin{table}[h]
\centering
\begin{tabular}{ccc}
\toprule
Module & Execution Time (ms) & Execution Time Proportion \\ \midrule
Ref. Path Generation & 4.69 & 9\% \\
Ref. Path Processing & 7.19 & 14\% \\
Path Optimization & 39.73 & 77\% \\ \bottomrule
\end{tabular}
\caption{Execution Time and Proportion for Each Module on CPU}
\label{tab:software_breakdown}
\end{table}

% \begin{table}[h]
% \centering
% \caption{\cmt{Execution Latency Breakdown of each Module on CPU}}
% \label{tab:software_breakdown}
% \resizebox{0.7\textwidth}{!}{%
% \begin{tabular}{cccc}
% \hline
%                                                                          & Ref. Path Generation & Ref. Path Process & Path Optimization \\ \hline
% \multirow{2}{*}{\begin{tabular}[c]{@{}c@{}}Avg. Latency \\ (ms)\end{tabular}} & 3.69                 & 5.19              & 41.73             \\
%                                                                          & 7.3\%                & 10.25\%           & 82.45\%           \\ \hline
% \end{tabular}%
% }
% \end{table}
\vspace{-1em}
\subsection{Design Space Exploration for Algorithm-Architecture Co-Optimization}

\subsubsection{Algorithm Optimization for faster convergence}
\label{sec:rho_opt}

\begin{figure}[ht]
    \centering

    \includegraphics[width=\linewidth]{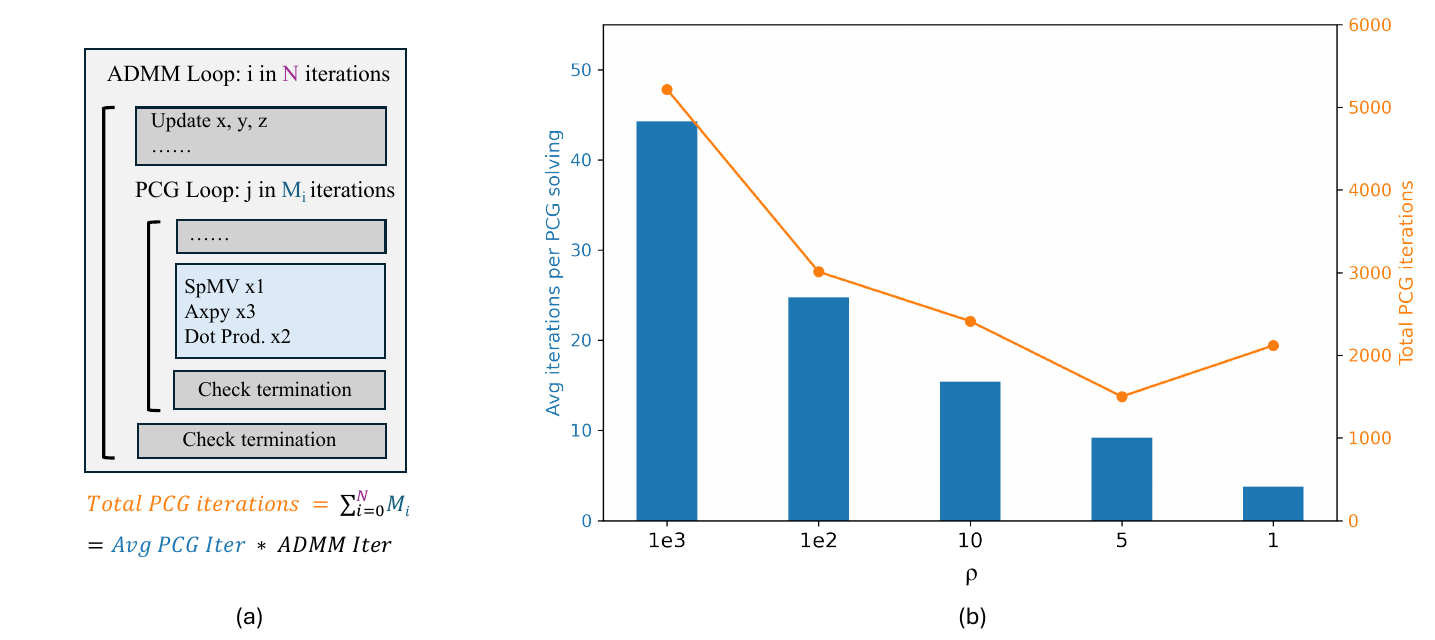}
    %\caption{The Structured Data Access for SpMV}
    \caption{Convergence Analysis. (a) Double nested loop structure in ADMM (b) Convergence speed on different $\rho$ setting}
    \Description{Convergence analysis }
    \label{fig:rho_op}
\end{figure}

In ADMM algorithm (Algorithm \ref{alg:ADMM}), the parameter $\rho$ represents the step-size, which can greatly affect the performance. \cite{osqp} use a large coefficient of $\rho$ to accelerate ADMM convergence. Our work uses an iterative method to solve the linear system, so the entire algorithm can be treated as a nested loop. Figure~\ref{fig:rho_op} (a) shows the loop structure of our algorithm. The inner PCG loop performs most of the calculations (in the blue box). However, experiment shows a large $\rho$ could significantly hinder the convergence of PCG. The opposite effect on inner and outer loop means that we cannot simply specify a value. To find the parameter setting corresponding to the global minimum of total PCG iterations, we performed a simulation-based search using a large number of real-collected data, the result is shown in Figure~\ref{fig:rho_op} (b). Although the PCG loop converges faster as $\rho$ decreases, the total number of iterations rebounds at value=1, indicating that value=5 best balances the inner and outer loops. Based on search results, we use an improved setting of $\rho$:

% In ADMM algorithm (Algorithm \ref{alg:ADMM}), a large $\rho$ for equation constraints can make ADMM converge faster~\cite{osqp}. However, our experiment shows an overly large $\rho$ could significantly hinder PCG convergence. Consider ADMM and PCG are both iterative methods, total PCG iterations = ADMM iterations  $\times$  Iterations in one PCG solving. Therefore, comparing the original setting in ~\cite{osqp}, we use an improved setting of $\rho$:

\begin{align}
\rho=diag(\rho_1,...,\rho_m), \quad \rho_i=
   \begin{cases}
		\bar{\rho} \quad  & l_i \neq u_i \\
		5\bar{\rho} \quad  &  l_i = u_i 
   \end{cases} \text{,}
   \label{eq:rho_i}
\end{align}
where $\bar{\rho}>0$ is initialized to 0.1. We also update $\bar{\rho}$ every 10 iterations to maintain the convergence of ADMM. 
% This new setting significantly reduces total PCG iterations. 

% \subsubsection{Low Bitwidth Optimization}
% \red{test}
% Additionally, we conducted tests on the impact of fixed-point numbers on precision, as shown in Table~\ref{tab:fixed}. It can be observed that Vitis HLS \texttt{ap\_fixed<24,9>} (24 total bits, 9 bits above decimal point) achieves the best accuracy. We observed that the input data for PCG consists of very small values, indicating a high requirement for decimal precision. Moreover, at least 8 bits are required to represent the integer part. Therefore, in the overall design, we ultimately opted for the fixed-point format  \texttt{ap\_fixed<24,9>}.

\subsubsection{Exploration of Mixed Precision Implementation}
The fixed-point arithmetic is widely adopted in many FPGA-based designs due to its lower resource usage and faster computation. In this work, we explore a float/fixed-point mixed precision scheme to optimize logic utilization while not compromising accuracy. For the Scaling module, the floating point format is necessary since it needs to deal with the excessive values in problem matrices. Then we only apply a 24-bit fixed-point format to the PCG module, because it has the most computational workload. We explore several combinations of 24-bit fixed-point format, with varying integer bits. We measure the numerical precision on more than 10 path planning samples that cover most scenarios. Table~\ref{tab:fixed} shows the precision for different data types. We count all the intermediate variables of PCG, and all the values are within the range of 9-bit integers. Therefore, we use ap\_fixed<24,9>, with 0.076 max error and 0.0065 mean error, indicating a centimeter-level error for 50-meter-level path planning. In the rest of the ADMM, we still use the floating-point format to minimize accuracy loss.

% \cmt{To thoroughly investigate the impact of fixed-point arithmetic on accuracy, we conducted a detailed exploration of mixed-precision implementation. We clarify that the selected 24-bit fixed-point format, \texttt{ap\_fixed<24,9>} (24 total bits, 9 bits for the integer part), is specifically applied only to the Preconditioned Conjugate Gradient (PCG) solver within the OSQP algorithm, rather than the scaling data step within the ADMM iterations. This selective application was motivated by preliminary observations indicating that ADMM iterations amplify arithmetic errors, potentially compromising solution accuracy. Consequently, the evaluation of fixed-point precision focused exclusively on the PCG solver by comparing its results directly against a floating-point baseline solution.}

% \cmt{We performed extensive profiling of intermediate variables across 10 diverse input scenarios to characterize data distribution comprehensively. 
% Figure~X illustrates the distribution of intermediate variables used in our analysis, highlighting the necessity of maintaining at least 8 integer bits to prevent overflow. 
% While exploring the fixed-point implementation, we initially assumed a default wrap-around arithmetic mode provided by Vitis HLS, deliberately avoiding the additional logic overhead required by saturated arithmetic. Under this configuration, overflow significantly impacted accuracy, underscoring the sensitivity of the PCG solver to arithmetic precision.}
Although the experiment results show the feasibility of the mixed-precision scheme, to ensure the robustness and generality across diverse application scenarios, even for possible extreme cases, we implement a fully floating-point version of the design.
The main challenge of full floating-point design is that floating-point arithmetic requires multiple clock cycles on FPGAs. In particular, the loop dependency of floating-point addition operations in the accumulation-based dot product operation (dot\_product+=a[i]*b[i]) will cause pipeline stalls. The full pipeline with Initiation Interval (II)=1 will be degraded to II=8, which greatly reduces performance. Our solution is to create 8 independent accumulation paths to break the loop-carrying dependency, each processing different data and merging the results at the end, reducing the II back to 1. 
The final full floating-point design achieves the same latency as the mixed-precision version, introducing moderate increases in logic resources. The detailed resource utilization comparison is shown in Section~\ref{sec:resource}. This allows users to choose between a full floating-point or a mixed-precision scheme according to their specific requirements and characteristics of the input data.

% \cmt{Furthermore, to address the impact of numerical outliers and ensure robustness and generality across diverse application scenarios, we also implemented a fully floating-point version of the design. This floating-point variant does not introduce additional clock cycles but incurs moderate increases in on-chip storage and additional DSP and register usage due to floating-point arithmetic operations. Detailed resource utilization comparisons presented in the experimental results section show approximately a 20\% increase in hardware resources for the fully floating-point design compared to the mixed floating-fixed precision version. Thus, the final implementation offers flexibility, allowing users to choose between fully floating-point precision or a hybrid floating-fixed precision according to the specific requirements and characteristics of their input data.}

% We also plotted the distribution of input data for PCG, as illustrated in Figure XXX, indicating a considerable demand for precision in the decimal places of the data.

\begin{table}[ht]
\centering
\caption{Comparison of Precision of Fixed-point Numbers}
\label{tab:fixed}
\begin{tabular}{cccc}
\toprule
data type                               & Max Error & MSE      & ME      \\ \midrule
ap\_fixed\textless{}20,9\textgreater{}  & 1.02      & 0.03     & 0.095    \\
ap\_fixed\textless{}24,8\textgreater{}  & 126.9     & 1403.8   & 27.195  \\
ap\_fixed\textless{}24,9\textgreater{}  & 0.076     & 0.000148 & 0.0065 \\
ap\_fixed\textless{}24,10\textgreater{} & 0.14      & 0.000577 & 0.013   \\
ap\_fixed\textless{}24,11\textgreater{} & 0.27      & 0.002    & 0.026   \\ \bottomrule
\end{tabular}
\vspace{-1em}
\end{table}

\section{Experimental Results}\label{sec:results}
\subsection{Experimental Setup}
\subsubsection{Hardware Platform and Tools}
We prototype and evaluate our design on the AMD ZCU102 Evaluation Kit, a Zynq UltraScale+ MPSoC embedded device equipped with a quad-core Arm Cortex-A53 with DDR memory (Processing System, PS), and FPGA fabric (Programming Logic, PL). We use High-Level Synthesis (HLS) to implement our design in AMD Vitis 2023.1 development tools. Both Scaling and ADMM IPs are synthesized and implemented on PL. The scaling module operates at 200MHz, with the ADMM module running at 250MHz.

\subsubsection{System-level Configurations}
PS and PL are interconnected via AXI interfaces, allowing PS-to-PL control and PL-DDR data movement. In this work, we use AXI4-Lite interface M\_AXI\_HPM0 for PS control and two AXI4 Memory Mapped interfaces S\_AXI\_HP\_\{0,1\} for transferring data. Each AXI4 Memory Mapped interface supports a maximum 128-bit width. Therefore, we pack four 32-bit elements into one data packet to maximize memory throughput. As we discussed in section~\ref{sec:systemlevel_pipeline}, for a complete path planning pipeline, the Reference Path Generation \& Processing steps are performed on CPU, then the pre-processed problem data for Path Optimization will be stored in DDR memory. The problem data includes objective matrix $P^{n\times n}$, constraint matrix $A^{m\times n}$, and constraint boundary $l^{m},u^{m}$ ($m = 1622$, $n=1619$). Since we implement pattern-aware matrix operations in this work, we only need to transfer the non-zero elements in matrix $P$ and $A$. Here,
we have $P_{NNZ}=(5L-1)$, $A_{NNZ}=(17L-5)$, $L=270$. Therefore, Path Optimization module need to transfer in total $(34L-2) = 9178$ elements from DDR Memory to PL. In order to accurately understand the impact of data movement on execution time, we perform an on-board test to profile the memory bandwidth. We transfer 1G bits of data from DDR to PL using two AXI4 interfaces and measure the transfer latency. The experiment result shows this configuration can achieve 7.2 GByte/s memory bandwidth, so we can transfer all optimization problem data within 4.75$\mu s$.

 %By integrating our pattern-specific design, we achieved significant speed improvements compared to CPU and ARM platforms. 
\subsubsection{Datasets and Algorithm Setup}

%Regarding the SpMV module, we compared our pattern-specific design with hard-coded SpMV design and traditional SpMV methods. The results show that our proposed SpMV design achieves a better balance between latency and resource utilization.
For evaluation, we use a public-available gridmap from an open-sourced path planning framework~\cite{pathoptimizergithub} as the input map. The gridmap has a size of 700$\times$700 pixels and 0.2m resolution, representing a 140m$\times$140m area with a series of obstacles. We sampled 40 path planning tasks in different areas of the map, divided them into four groups according to their difficulty level \{\textit{Easy}, \textit{Medium 1}, \textit{Medium 2}, \textit{Hard}\}. Since the ADMM algorithm (Algo.~\ref{alg:ADMM}) for QP solving is iterative, to ensure the accuracy and convergence of the QP solution, we use the same hyperparameter setting in OSQP~\cite{osqp}, including $\sigma$ and $\alpha$. We also apply the same residual tolerance for convergence checking. For parameter $\rho$, we use the optimized setting in section~\ref{sec:rho_opt}.

% Firstly, we presented the planning results of our designed planner for different paths in Section~\ref{sec:planning_result}. Considering that this design is pattern customized, we compared our pattern-specific design with hard-coded SpMV design and traditional SpMV methods in Section~\ref{sec:spmv}. To reduce solution latency, Section~\ref{sec:optimization} presents the ablation study on all proposed optimizations. In Section~\ref{sec:time_cons}, we have compared other QP solvers with our design. Finally, we elucidated our hardware resource utilization in Section~\ref{sec:resource}. 

\subsection{Planning Results}
\label{sec:planning_result}

\begin{figure}[h]
    \centering
\includegraphics[width=\linewidth]{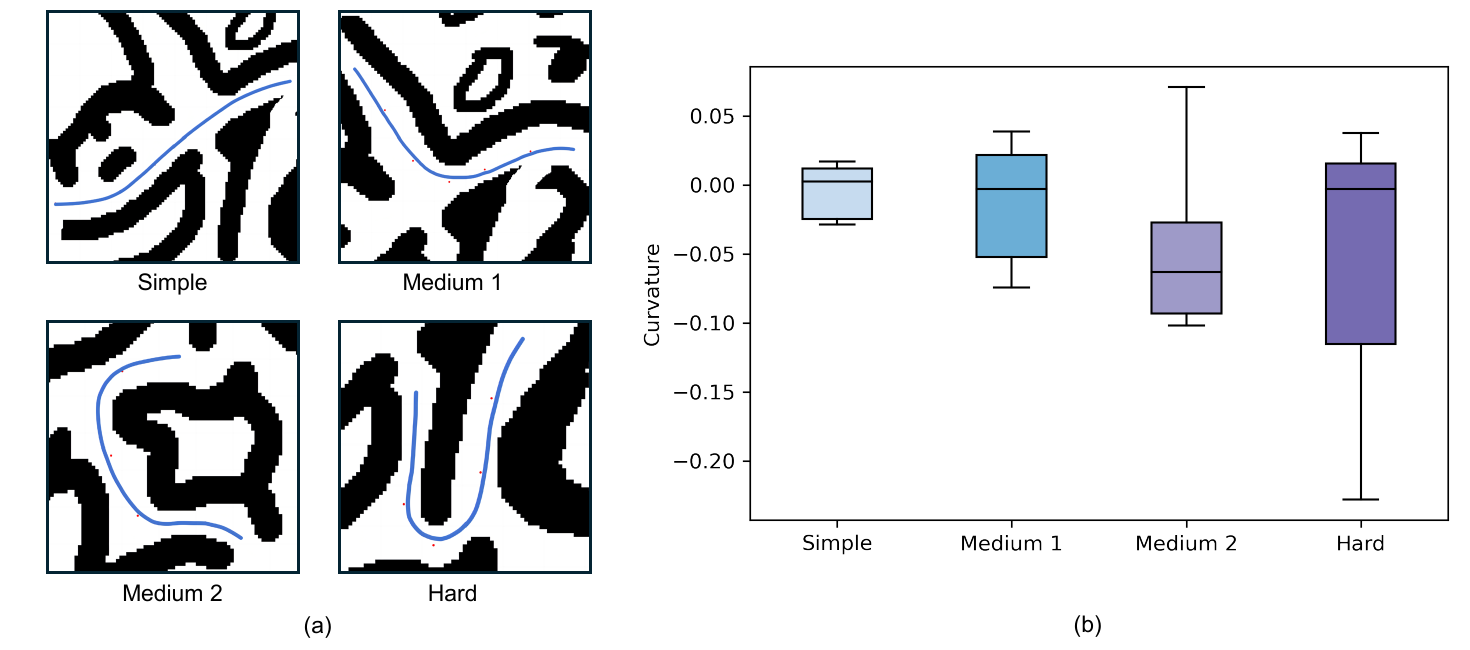} 

    \caption{Planning results of four representative paths. (a) visualization of the optimized path (b) curvature distribution }
    \Description{This figure displays the planning results under different paths. }
\label{fig:path1-4}
\vspace{-1em}
\end{figure}

We evaluate our proposed path planning framework on four groups of paths of varying difficulty levels. The primary factor influencing planning difficulty is curvature. We arrange these four groups of paths in order of increasing difficulty (\textit{Simple}, \textit{Medium 1}, \textit{Medium 2}, \textit{Hard}), exhibiting distinct curvature distributions. Four planning results and the curvature distributions are visualized in Figure~\ref{fig:path1-4}. Path \textit{Simple} is a smooth path with the smallest absolute curvature values and the narrowest range. The path \textit{Hard} demonstrates a U-turn in a narrow space. It displays relatively large overall absolute curvature values, with a wider distribution span and, therefore, the hardest planning difficulty. We then evaluate our work on these four representative paths.

% \begin{figure}
%     \centering
%     \includegraphics[width=0.5\linewidth]{Figures/k_distribution.pdf}
%     \caption{Curvature Distribution of Four Paths}
%     \label{fig:k_distribution}
%     %% \vspace{-1em}
% \end{figure}

%\begin{table}
%    \centering
%    \caption{Comparison of Delays for 10 ADMM Iterations under Different Optimization Designs}
%    \label{tab:optimizetion}
%    \begin{tabular}{cccc}
%    \toprule
%     Unroll Factor& Parallelly Compute    & Parallelization   & Latency  \\
%     for Vectors & $x$, $r$, $y$ in PCG & Vectors Update & (ms)\\
%    \midrule
%       6 & No & No  & 1.59 \\
%       12 & No  & No & 1.25\\
%       12 & Yes  & No & 0.906\\
%       \textbf{12} & \textbf{Yes} & \textbf{Yes} & \textbf{0.882}\\
%    \bottomrule
%    \end{tabular}
%\end{table}

\subsection{Comparison with Existing Works}
\label{sec:time_cons}
% In this study, we conducted a comprehensive performance evaluation of our hardware-designed solver on different hardware platforms. Our assessment encompassed two key platforms: the Intel i5 CPU and ARM architecture. The outcomes vividly depict the exceptional speed performance of our solver on both of these platforms. In contrast to conventional CPUs, our solver achieved a remarkable to-be-write acceleration, showcasing a substantial boost in speed. Similarly, when compared to the ARM platform, we achieved an equally significant to-be-write acceleration. These results underscore the versatility of our hardware design solver, highlighting its ability to excel in diverse hardware environments and catalyze efficient solution technology applications across a spectrum of domains. The implications of these findings are substantial, as they contribute to the enhancement and optimization of performance research and practical applications. Table \ref{tab:com CPU ARM} shows the comparison of time consumption.

\begin{figure}[]
    \centering

    \includegraphics[width=\linewidth]{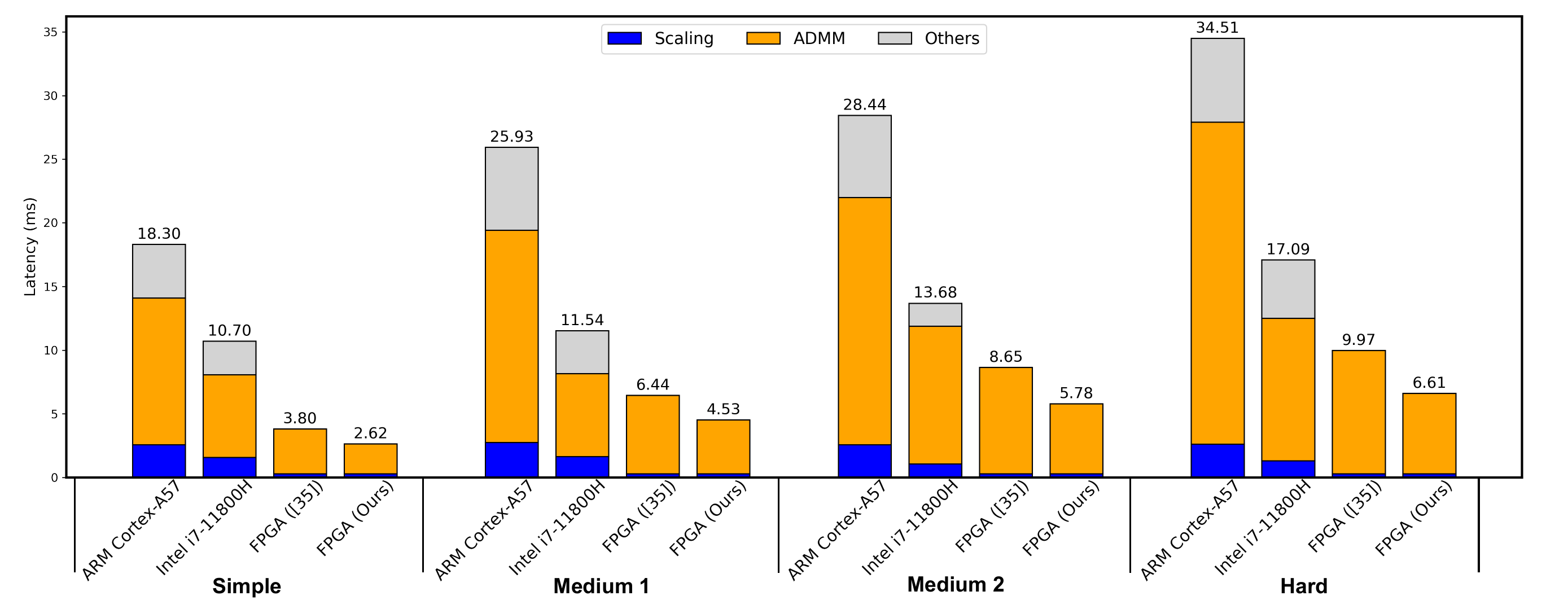}
    \caption{Comparison of computation latency}
    \Description{This figure shows the breakdown of execution time. }
    \label{fig:time_percent}
\end{figure}

% In this section, we evaluated the performance of the state-of-art QP solving framework OSQP~\cite{osqp} on various platforms like CPU and GPU, as well as our proposed FPGA-based design. In addition, we also compare other FPGA-based QP solver with ours. 
\subsubsection{Comparison of QP solving time}
We first compare our work with the state-of-the-art FPGA-based design \cite{pathplanner_iccad}, and the OSQP~\cite{osqp} solver on the Intel i7-11800H and ARM Cortex-A57 CPU, since OSQP is the state-of-the-art QP solving framework dedicated to the CPUs. We measure the computation latency of QP solving on four representative paths mentioned in section~\ref{sec:planning_result}. As Figure~\ref{fig:time_percent} shows, our design demonstrates a significant performance improvement over existing works, including an average 1.48 $\times$ speedup over state-of-the-art FPGA-based work, a 2.89 $\times$ speedup over the Intel CPU on average. Compared to ARM, it achieves a 5.62$\times$ speed improvement on average. In the evaluation, the Intel CPU runs at 2.3GHz and ARM runs at 1.43GHz.

% While most existing QP solvers run on CPU, there has been recent interest in GPU due to its massive parallelism to accelerate solutions of QPs. 
% \cite{GPUAcc} proposes cuOSQP, a GPU-based quadratic programming solver that achieves speed-ups over OSQP on large-scale problems. We test cuQSOP on NVIDIA RTX 3090 GPU with same problem settings. Table~\ref{tab:gpu_result} compares the performance of our design with cuOSQP. Our specialized architecture demonstrates over 18.1$\times$ speed-up over GPU-based implementation.
% A possible explanation for this large performance gap is that GPU cannot fully utilize the computing units to amortize kernel launch and data movement overhead in iterative solving.
% In contrast, our pattern-aware design fully utilizes computing units for parallelized operations and minimizes data movement.

In Table~\ref{tab:comp_all}, we comprehensively compare the characteristics of existing works, including linear solver type, optimization approaches, problem size (constraints and variables), latency, and power consumption. Our design leverages unique data flow optimizations and system-level pipelining.

\begin{table}[]
\centering
\resizebox{0.95\textwidth}{!}{%
\begin{tabular}{|c|cccccc|}
\hline
 &
  \multicolumn{1}{c|}{\textbf{This Work}} &
  \multicolumn{1}{c|}{\cite{pathplanner_iccad}} &
  \multicolumn{2}{c|}{OSQP\cite{osqp}} &
  \multicolumn{1}{c|}{cuOSQP\cite{GPUAcc}} &
  ReLU-QP\cite{relu-qp} \\ \hline
Platform &
  \multicolumn{1}{c|}{FPGA} &
  \multicolumn{1}{c|}{FPGA} &
  \multicolumn{1}{c|}{CPU} &
  \multicolumn{1}{c|}{CPU} &
  \multicolumn{1}{c|}{GPU} &
  GPU  \\ \hline
Arch. &
  \multicolumn{1}{c|}{ZCU102} &
  \multicolumn{1}{c|}{ZCU102} &
  \multicolumn{1}{c|}{Intel i7-11800H} &
  \multicolumn{1}{c|}{ARM Cortex-A57} &
  \multicolumn{1}{c|}{RTX 3090} &
  RTX 3090 Ti \\ \hline
Linear Solver &
  \multicolumn{1}{c|}{PCG} &
  \multicolumn{1}{c|}{PCG} &
  \multicolumn{1}{c|}{LDLt} &
  \multicolumn{1}{c|}{LDLt} &
  \multicolumn{1}{c|}{PCG} &
  N/A  \\ \hline
Sparsity Utilization &
  \multicolumn{1}{c|}{Customized Arch.} &
  \multicolumn{1}{c|}{Customized Arch.} &
  \multicolumn{1}{c|}{General Algo.} &
  \multicolumn{1}{c|}{General Algo.} &
  \multicolumn{1}{c|}{CUDA lib} &
  Torch lib \\ \hline
Dataflow Optimization &
  \multicolumn{1}{c|}{\Checkmark} &
  \multicolumn{1}{c|}{\XSolidBrush} &
  \multicolumn{1}{c|}{\XSolidBrush} &
  \multicolumn{1}{c|}{\XSolidBrush} &
  \multicolumn{1}{c|}{\XSolidBrush} &
  Torch.jit \\ \hline
\multicolumn{1}{|l|}{Sample-wise Pipeline} &
  \multicolumn{1}{c|}{\Checkmark} &
  \multicolumn{1}{c|}{\XSolidBrush} &
  \multicolumn{1}{c|}{\XSolidBrush} &
  \multicolumn{1}{c|}{\XSolidBrush} &
  \multicolumn{1}{c|}{\XSolidBrush} &
  \XSolidBrush \\ \hline
Constraints \# &
  \multicolumn{6}{c|}{1622}\\ \hline
Variables \# &
  \multicolumn{6}{c|}{1619} \\ \hline
Latency (ms) &
  \multicolumn{1}{c|}{\textbf{4.87}} &
  \multicolumn{1}{c|}{7.19} &
  \multicolumn{1}{c|}{13.25} &
  \multicolumn{1}{c|}{26.81} &
  \multicolumn{1}{c|}{88.2} &
  7.63 \\ \hline
Power (W) &
  \multicolumn{1}{c|}{10.7} &
  \multicolumn{1}{c|}{10.29} &
  \multicolumn{1}{c|}{35} &
  \multicolumn{1}{c|}{3.4} &
  \multicolumn{1}{c|}{224} &
  281 \\ \hline
\begin{tabular}[c]{@{}c@{}}Energy Efficiency \\ (Samples/J)\end{tabular} &
  \multicolumn{1}{c|}{\textbf{19.2}} &
  \multicolumn{1}{c|}{13.5} &
  \multicolumn{1}{c|}{2.15} &
  \multicolumn{1}{c|}{10.97} &
  \multicolumn{1}{c|}{0.05} &
  0.47 \\ \hline
\end{tabular}%
}
\caption{Comprehensive Comparison with Existing Works}
\label{tab:comp_all}
\vspace{-2em}
\end{table}

We also compare our work with GPU-based works.
While most existing QP solvers run on CPU, there has been recent interest in GPU due to its massive parallelism to accelerate solutions of QPs. 
\cite{GPUAcc} proposes cuOSQP, a GPU-based quadratic programming solver that achieves speed-ups over OSQP on large-scale problems. We test open-sourced cuQSOP on NVIDIA RTX 3090 GPU with same problem settings. 
In the host program, the problem data is first stored in the main memory and then transferred to the GPU, and the latency is measured with the built-in timer. 
In table~\ref{tab:comp_all}, we show the average latency on cuOSQP. Our specialized architecture demonstrates over 18.1$\times$ speed-up over GPU-based implementation.
A possible explanation for this large performance gap is that GPU cannot fully utilize the computing units to amortize kernel launch and data movement overhead in iterative solving.
In contrast, our pattern-aware design fully utilizes computing units for parallelized operations and minimizes data movement.

 \cite{relu-qp} proposes ReLU-QP, a GPU-accelerated quadratic programming solver that reformulates sequential ADMM updates into a weight-tied deep neural network with ReLU activations, enabling the deployment using standard machine-learning toolboxes like PyTorch, which features highly optimized computing kernels on GPUs. We evaluated the open-sourced ReLU-QP project on the same path-planning dataset using an RTX 3090 Ti GPU. Note that the ReLU-QP host program divides the total execution time into "setup time" and "solve time", with host-GPU data transfers and KKT matrix computations attributed to the "setup time", which can take several seconds. At the same time, our implementation includes data transfer and matrix computations within the reported execution times. In table~\ref{tab:comp_all}, we only include the "solve time," yet our approach still achieves a 1.56$\times$ speedup and a 40$\times$ improvement in energy efficiency.

 \subsubsection{Comparison with other FPGA-based works}
 % \cite{jerez2011qpfpga} implements a QP solver on FPGA, which also utilizes the sparsity of the problem. Our multi-level optimized design demonstrates a significant advancement. Our work solves the problem with 1622 constraints and 1619 variables in 4.87 ms on average, while \cite{jerez2011qpfpga} costs 371 ms to solves the problem with 560 constraints and 416 variables. 
We compared our work with other FPGA-based QP solver on model predictive control (MPC) problems of similar scale (measured by \#non-zero elements of matrix A and P). RSQP~\cite{RSQP} is a general-purpose QP solver that enables Problem-specific customization on hardware. RSQP offloads the PCG solving in the ADMM method to an FPGA, and encodes the matrix sparse patterns for efficient computing.
RSQP is implemented on the data-center-grade AMD U50 FPGA with HBM for data movement. Such an approach demands significantly more hardware resources, making it unsuitable for edge scenarios. Besides, due to RSQP's heterogeneous architecture, it involves frequent CPU-FPGA communication. In each ADMM iteration, the solution vector needs to be transferred back to CPU to perform the vector update.
In contrast, our solver is fully FPGA-based, implemented entirely on an embedded ZCU102 platform. Our work only requires problem data transaction once, eliminating the frequent CPU-FPGA communication overhead in RSQP, leading to a 40$\times$ speedup and 100$\times$ improvement in energy efficiency.
In the state-of-the-art general FPGA-based QP solver~\cite{fpgaqp_micro2024}, it proposes OSQP-direct and OSQP-indirect, which accelerate the linear system solutions in the ADMM algorithm by exploiting problem-specific sparsity patterns. Specifically, it introduces a pipelined spatial architecture called Multi-Issue Butterfly (MIB) that efficiently schedules scalar, vector, and matrix operations based on these sparsity patterns. For a fair comparison, we evaluate against its PCG-based QSOP-indirect implementation. With the highly customized sparsity pattern optimization and fully pipelined architecture tailored to our path-planning scenario, our approach achieves a 2.67× speedup and a 6.8× improvement in energy efficiency.
We further compare our PCG solving performance with Callipepla~\cite{Callipepla}, a dedicated PCG solver on an FPGA. Callipepla proposes a stream-centric architecture and leverages a general SpMV unit to reduce latency. The HBM-equipped U280 data-center FPGA also enables massive parallelism and extremely high memory bandwidth. For fair comparison, we compare the non-zero elements processing throughput, which normalizes the performance for sparse matrices with different scales. Table~\ref{tab:Callipepla} shows our highly customized architecture on an embedded FPGA with multi-level dataflow optimization achieves 1.09$\times$ speedup and 7.14$\times$ energy efficiency, with significantly fewer hardware resources, which demonstrates the importance and effectiveness of domain-specific customization on resource-constrained platforms.
\begin{table}[]
\centering
\caption{Comparison with State-of-the-art general QP solver on FPGA}
\label{tab:qp_fpga}
{%
\begin{tabular}{cccc}
\toprule
                            & \textbf{This Work}  & 
                        \textbf{OSQP-indirect\cite{fpgaqp_micro2024}} & \textbf{RSQP\cite{RSQP} }        \\ \midrule
Linear Solver               & PCG        & PCG                     & PCG          \\
nnz(P)+nnz(A)               & 5934       & 5880                    & 5880         \\
\textbf{Solve Time (ms)}             & \textbf{4.87}       & 13                      & 195          \\
FPGA Platform                    & AMD ZCU102 & AMD U50                 & AMD U50      \\
Frequency (MHz)             & 250        & 236                     & 236          \\
Memory Bandwidth            & 7.2        & 57.6                    & 28.8 - 115.2 \\
DSP                         & 674        & 952                     & N/A          \\
LUTs                        & 146K       & 279K                    & N/A          \\
Power (W)                   & 7          & 18                      & 19           \\
\textbf{Energy Efficiency (\#Solution/s/W)} & \textbf{29.3 }      & 4.3                     & 0.27         \\ \bottomrule
\end{tabular}%
}
\end{table}

\subsubsection{Comparison of end-to-end throughput}
\begin{figure}[h]
    \centering

    \includegraphics[width=\linewidth]{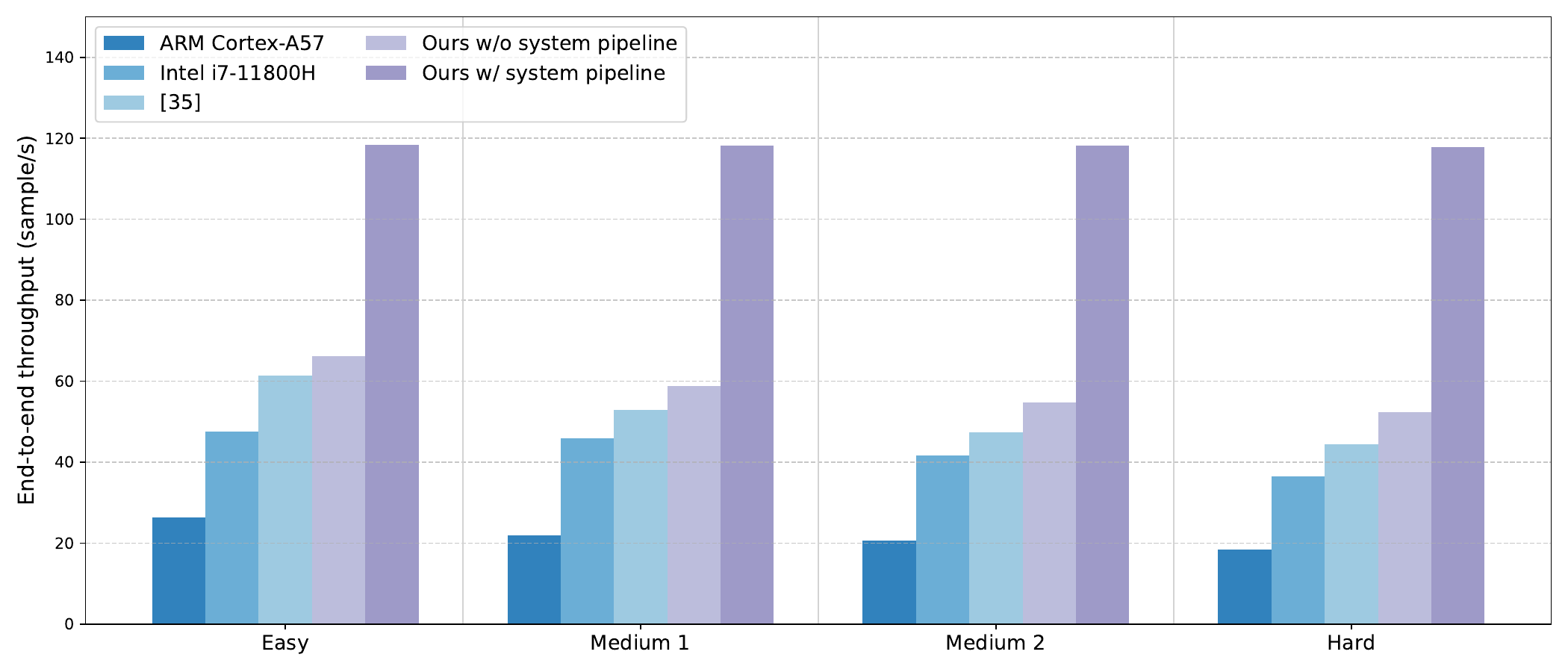}
    \caption{Comparsion of end-to-end path planning throughput}
    \label{fig:end2end_throughput}
\end{figure}

We further perform the evaluation on path planning end-to-end throughput to demonstrate the impact of our system-level optimization. The results are shown in Figure~\ref{fig:end2end_throughput}. In our design, the system pipeline achieves 2$\times$ end-to-end throughput improvement. Compared with other existing works, our work achieves 2.4$\times$-5.9$\times$ end-to-end throughput improvement. 

% Please add the following required packages to your document preamble:
% \usepackage{graphicx}
\begin{table}[]
\centering
\caption{Comparision with State-of-the-art FPGA-based PCG Solver}
\label{tab:Callipepla}
\resizebox{0.9\textwidth}{!}{%
\begin{tabular}{ccccccccc}
\hline
 &
  \multicolumn{3}{c}{\textbf{Problem Setting}} &
  \multicolumn{2}{c}{\textbf{Performace}} &
  \multicolumn{3}{c}{\textbf{Resource Untilization}} \\ \hline
 &
  \#Row &
  NNZ &
  \multicolumn{1}{c|}{Sparsity} &
  \begin{tabular}[c]{@{}c@{}}Throughput$^\dagger$\\ (GFLOP/s)\end{tabular} &
  \multicolumn{1}{c|}{\begin{tabular}[c]{@{}c@{}}Energy Efficiency\\ (GFLOP/s/W)\end{tabular}} &
  Hardware &
  LUTs &
  DSP \\ \hline
Callipepla\cite{Callipepla} &
  3,948 &
  117,816 &
  \multicolumn{1}{c|}{99.2\%} &
  18.84 &
  \multicolumn{1}{c|}{0.41} &
  \begin{tabular}[c]{@{}c@{}}U280 \\ Data Center FPGA\end{tabular} &
  509k &
  1940 \\ \hline
\textbf{This Work} &
  1,619 &
  9,666 &
  \multicolumn{1}{c|}{99.6\%} &
  \begin{tabular}[c]{@{}c@{}}20.51\\ (1.09$\times$)$^\ddagger$\end{tabular} &
  \multicolumn{1}{c|}{\begin{tabular}[c]{@{}c@{}}2.92 \\ (7.14$\times$)\end{tabular}} &
  \begin{tabular}[c]{@{}c@{}}ZCU102 \\ Embedded FPGA\end{tabular} &
  146k &
  674 \\ \hline
\end{tabular}%
}
$^\dagger$Throughput: \#Non-zero elements per second \hspace{1em} $^\ddagger$Improvement compared with Callipepla
\end{table}

% \subsubsection{Comparison with other FPGA-based QP solver}

% \cite{pathplanner_iccad}

\subsection{Ablation Study on Proposed Optimization Approaches}
\label{sec:optimization}

\subsubsection{Impact of Sparse Matrix-Vector Multiplication Unit}

To demonstrate the effectiveness of proposed optimizations, we perform a comprehensive ablation study. We first evaluate the impact of the proposed pattern-aware SpMV. The QP problem setup leads to a structural sparsity pattern of the problem matrices, as discussed in Section~\ref{sec:problem_matrix}. A naive idea to leverage this pattern is to hard-code it into the HLS code. We developed a code generator that takes the matrix as input and generates HLS kernel code with the pattern hard-coded. As Table~\ref{tab:SpMV} shows, the SpMV alone consumes >50\% of the total logic resources, which is unacceptable. 
\begin{table}[h]
    \centering
    %\vspace{-1em}
    \vspace{-0.5em}
    \caption{Comparison of Different Implementations of SpMV}
    \vspace{-1em}
    \label{tab:SpMV}
    \begin{tabular}{ccccc}
    \toprule
     \multirow{2}{*}{SpMV Versions}& Latency & \multirow{2}{*}{LUT} & \multirow{2}{*}{Flip-Flops} & \multirow{2}{*}{DSP} \\
     \multirow{2}{*}{} & (clock cycles) &\multirow{2}{*}{} &\multirow{2}{*}{} & \multirow{2}{*}{}\\
    \midrule
       \multirow{2}{*}{General CSC-based} & \multirow{2}{*}{1621}  & 5841 & 2998  & 16 \\
       \multirow{2}{*}{} & \multirow{2}{*}{} & (2\%) & (\textasciitilde{}0\%) & (\textasciitilde{}0\%) \\ \hline
       \multirow{2}{*}{Pattern Hard-coded} & \multirow{2}{*}{546} & 156051 & 338009 & 32 \\
       \multirow{2}{*}{} & \multirow{2}{*}{} & (56\%) & (61\%) & (1\%) \\ \hline
       \multirow{2}{*}{Our Proposed} & \multirow{2}{*}{279} & 4078 & 8767 & 102 \\
       \multirow{2}{*}{} & \multirow{2}{*}{} & (1\%) & (1\%) & (4\%) \\
    \bottomrule
    \end{tabular}
    \vspace{-1em}
\end{table}

Besides, since we use CSC (Compressed Sparse Column) format for the sparse matrices. We use a general CSC-based SpMV Unit as the baseline, which uses column pointers and row indices to access the non-zero values. We compare the baseline with our proposed pattern-aware SpMV unit. As shown in Table~\ref{tab:SpMV}, our SpMV unit saves ~1200 clock cycles with <4\% additional logic resources. From Section~\ref{sec:rho_opt}, we know that a QP solving typically requires >1000 PCG iterations. Therefore, the proposed SpMV unit can reduce at least 4.8ms latency.

\subsubsection{Impact of Multi-Level Optimization}
% In our design, the ADMM module reflects higher latency, and we have gradually optimized the ADMM design. 

We further evaluate the proposed multi-level optimization in Section~\ref{sec:dataflow_opt}. As discussed in section~\ref{sec:spmv}, our pattern-aware SpMV unit can generate 6 output elements per clock cycle. Besides SpMV, the PCG solving involves several vector operations (two dot products, three AXPY). Failing to consider these operators will significantly increase the latency. We implement parallelization for these vector operators, with factor=6 (6 outputs/cycle), as the baseline. After parallelization, the latency of all operators is reduced to \#Points $L$. Then we increase the parallel factor to 12 by implementing more logic units. To perform the ablation study, we incrementally evaluate each optimization on 10 path planning samples with varying difficulty. Table~\ref{tab:ablation} shows the average latency with resource utilization for each optimization combination. We set the design with parallel factor=6 as the baseline. Increasing the parallel factor to 12 reduces 31\% latency; however, it also introduces significantly more logic resources. The algorithm parameter optimization reduces 28\% with no additional overhead. Finally, the dataflow optimization further reduces 42\% latency. The resource overhead is from implementing the inter-operator pipeline using extra registers and interconnections. This demonstrates the effectiveness and efficiency of the proposed optimizations.

% To demonstrate the effectiveness of proposed optimizations, we perform an ablation study, as shown in Figure~\ref{fig:}. We first evaluate the performance with the sparse-aware parallelization-only scheme (section~\ref{sec:sparse_opt}) with unroll factor = 6, i.e. 6 elements output per clock cycle. 
% Then we increase the unroll factor to 12. We further introduce the algorithm parameter optimization discussed in section~\ref{sec:rho_opt}.
% Finally, we introduce the proposed multi-level dataflow optimization in \ref{sec:dataflow_opt}. The result shows that the dataflow optimization has the biggest impact on performance, achieves 70\% improvement against previous optimizations.

\begin{table}[]
\centering
\caption{Ablation study on proposed optimization approaches}
\label{tab:ablation}
\resizebox{0.6\columnwidth}{!}{%
\begin{tabular}{c|cccc}
\hline
                  &  \textcircled{1}     & \textcircled{2}      & \textcircled{2}+\textcircled{3} & \textcircled{2}+\textcircled{3}+\textcircled{4} \\ \hline
Avg. Latency (ms) & 14.3    & -4.5 (-31\%) & -2.7 (-28\%)  & -3 (-42\%)    \\ 
LUT               & 46326 & +21177  & +0  & +2734  \\ 
DSP               & 295   & +56    & +0  & +26    \\
BRAM              & 103   & +32    & +0   & +10      \\
FF               & 40955  & +31150  & +0   & +6744     \\
\hline
\end{tabular}%
}\\
\textcircled{1} Parallel Factor=6 \quad \quad \textcircled{2}Parallel Factor=12\\
\textcircled{3} Parameter Optimization (Sec.~\ref{sec:rho_opt}) \textcircled{4} Dataflow Optimization (Sec.~\ref{sec:dataflow_opt})
\end{table}

\subsection{Resource Utilization}
\label{sec:resource}

The hardware resource consumption of our implementation is shown in Table~\ref{tab:Resource utilizationt}. 
% We have utilized a significant number of LUT resources in the PCG method. The PCG method is implemented within the ADMM module, leading to a LUT resource utilization of 21\%.
The final hardware implementation of OSQP used 26.8\% DSP, 53.3\% LUT, 25.4\% BRAM, and 37.1\% registers of ZCU102.
Compared with the standard mixed-precision implementation, the full floating-point version introduces 13\% more DSP, 62\% more BRAM, as well as slightly more LUT and FF.
%As solving linear equations within the ADMM is a critical point for hardware acceleration, the PCG method maximizes the parallel computing advantages of FPGA.
% \begin{table}[ht]
%     \centering
%     % \vspace{-0.5em}
%     \caption{FPGA Hardware Resource Consumption}
%     % \vspace{-1em}
%     \label{tab:Resource utilizationt}
%     \begin{tabular}{ccccc}
%     \toprule
%      Module& LUT & DSP & BRAM & Register\\
%     \midrule
%         \multirow{2}{*}{Scaling} & 76001  &  297 & 87 & 124949\\
%         \multirow{2}{*}{} &  (27.73\%) & (11.79\%) & (9.54\%)  &  (22.79\%) \\ \hline
%         \multirow{2}{*}{ADMM} & 113754  &  400 & 53.5 & 92780\\
%         \multirow{2}{*}{} & (41.50\%) & (15.87\%) & (5.87\%) & 16.93\%) \\
%     \midrule
%         \multirow{2}{*}{\textbf{Total}} & 194780  &  697 & 140.5 & 224588 \\
%         \multirow{2}{*}{} & (71.07\%)  &  (27.66\%) & (15.41\%) & (40.97\%) \\
%     \bottomrule
%     \end{tabular}
%     % \vspace{-2em}
% \end{table}

% Please add the following required packages to your document preamble:

\begin{table}[h!]
    \centering
    \caption{FPGA Hardware Resource Consumption}
    \label{tab:Resource utilizationt}
\begin{tabular}{|c|cccc|}
\hline
Module                          & LUT      & DSP      & BRAM      & FF        \\ \hline
Scaling                         & 76001    & 297      & 87        & 124949    \\ \hline
ADMM                            & 71237    & 377      & 145       & 78849     \\ \hline
\multirow{2}{*}{\textbf{Total}} & 147238   & 674      & 232       & 203798    \\
                                & (53.6\%)   & (26.8\%)  & (25.4\%)  & (37.1\%)    \\ \hline
\multirow{2}{*}{\textbf{Total (float)}}       & 151635   & 762      & 377     & 217495    \\
                                & (55.2\%) & (30.3\%) & (41.3\%) & (39.6\%) \\ \hline
\end{tabular}
\end{table}
\section{Conclusion}\label{sec:conclusion}
Most commercial autonomous vehicles rely on computationally intensive path planners, which place heavy demands on computation platforms. To address this, we proposed a novel, sparsity-aware FPGA-based path planning approach with  HW/SW co-design. By exploiting structural patterns in the problem matrix, we designed an efficient storage scheme and processing units. 
With our multi-level dataflow optimization, our work achieves superior performance over state-of-the-art implementations while balancing computation time and resource usage.

\bibliographystyle{unsrt}
\bibliography{references}

\end{document}